\documentclass[a4paper,12pt]{article}
\pdfoutput=1
\usepackage{epsfig,graphicx,xcolor,amsbsy,amssymb,latexsym,amsfonts,amsmath,,tcolorbox,setspace}
\usepackage{pstricks}
\usepackage{color}
\usepackage{soul}
\usepackage[numbers,square,comma, compress]{natbib}
\usepackage{placeins}
\usepackage{wrapfig,framed,caption}

\definecolor{shadecolor}{gray}{0.925}

\usepackage{eurosym}
\usepackage[a4paper]{geometry}
\usepackage{graphicx}
\usepackage{tikz}
\usepackage{xcolor}
\usepackage{amsmath,amssymb,amsfonts,pstricks,setspace}
\usepackage[enableskew]{youngtab}

\numberwithin{equation}{section}

\newcommand{\bea}{\begin{eqnarray}\displaystyle}
\newcommand{\eea}{\end{eqnarray}}

\newcommand{\figref}[1]{Fig.~\protect\ref{#1}}

\newcommand{\propOrb}{\mathcal{O}^{(2),0}}
\newcommand{\propNonOrb}{\mathcal{O}^{(2),1}}

\newcommand{\cpf}[2]{\mathcal{O}^{(#1),#2}}

\title{
\begin{flushright}{\vspace{-2.5cm}\small LYCEN 2019-03\\}\end{flushright}
\vspace{2.3cm}
{\bf From Little String Free Energies\\ Towards Modular Graph Functions}\\[40pt]}

\author{\large \textsc{Stefan~Hohenegger\footnote{\tt s.hohenegger@ipnl.in2p3.fr}}}
\date{}

\begin{document}

\maketitle
\thispagestyle{empty}
\begin{center}
\renewcommand{\thefootnote}{\fnsymbol{footnote}}\vspace{-0.5cm}
${}^{\footnotemark[1]}$ Univ Lyon, Univ Claude Bernard Lyon 1, CNRS/IN2P3, IP2I Lyon,\\ UMR 5822, F-69622, Villeurbanne, France\\[2.5cm]
\end{center}

\begin{abstract}
We study the structure of the non-perturbative free energy of a one-parameter class of little string theories (LSTs) of A-type in the so-called unrefined limit. These theories are engineered by $N$ M5-branes probing a transverse flat space. By analysing a number of examples, we observe a pattern which suggests to write the free energy in a fashion that resembles a decomposition into higher-point functions which can be presented in a graphical way reminiscent of sums of (effective) Feynman diagrams: to leading order in the instanton parameter of the LST, the $N$ external states are given either by the fundamental building blocks of the theory with $N=1$, or the function that governs the counting of BPS states of a single M5-brane coupling to one M2-brane on either side. These states are connected via an effective coupling function which encodes the details of the gauge algebra of the LST and which in its simplest (non-trivial) form is captured by the scalar Greens function on the torus. More complicated incarnations of this function show certain similarities with so-called modular graph functions, which have appeared in the study of Feynman amplitudes in string- and field theory. Finally, similar structures continue to exist at higher instanton orders, which, however, also contain contributions that can be understood as the action of (Hecke) operators on the leading instanton result. 
\end{abstract}

\newpage

\tableofcontents

\onehalfspacing

\vskip1cm

\section{Introduction}
In a series of works \cite{Hohenegger:2016yuv,Bastian:2017ing,Bastian:2017ary,Bastian:2018dfu} it has been argued that a certain class of $\mathcal{N}=1$ supersymmetric \emph{little string theories} (LSTs) \cite{LSTR1,LSTR2,LSTR3,LSTR4,LSTR4.5,LSTR5,LSTR6,LSTR7,Bhardwaj:2015oru,Bhardwaj:2019hhd,Haouzi:2017vec} of A-type \cite{Hohenegger:2015btj,Hohenegger:2016eqy} exhibits an intricate web of dualities. These theories are engineered in M-theory by $N$ parallel M5-branes spread out on a circle $S^1$ and probing a transverse $\mathbb{Z}_M$ geometry. A more geometric description of these theories can be given in terms of F-theory compactified on a particular type of toric Calabi-Yau threefold $X_{N,M}$ \cite{Kanazawa:2016tnt} which exhibits the structure of a double elliptic fibration. 

While these LSTs contain string-like degrees of freedom (albeit without gravity), their low energy limits are supersymmetric gauge theories in six dimensions. It was argued in \cite{Bastian:2017ary} that, for given $(N,M)$, the Calabi-Yau geometry $X_{N,M}$ engineers three (generically) different supersymmetric gauge theories, a fact which was termed \emph{triality}. These theories are very different regarding their matter- and gauge content and are related through intrinsically non-perturbative duality maps, which mixes coupling constants, Coulomb branch- and gauge parameters.

Furthermore, it was conjectured in \cite{Hohenegger:2016yuv} (and shown explicitly in a large number of examples) that the Calabi-Yau threefolds $X_{N,M}$ and $X_{N',M'}$ are dual\footnote{The duality $X_{N,M}\sim X_{N',M'}$ means that these two manifolds share the same extended K\"ahler moduli space, \emph{i.e.} their K\"ahler moduli spaces are connected through so-called flop transformations. We refer to \cite{Hohenegger:2016yuv} for details.} to one another if $NM=N'M'$ and $\text{gcd}(N,M)=\text{gcd}(N',M')$. It is therefore expected that the (non-perturbative) BPS-partition functions $\mathcal{Z}_{N,M}(\omega,\epsilon_{1,2})$ and $\mathcal{Z}_{N',M'}(\omega',\epsilon_{1,2})$ agree once a suitable duality map (conjectured in \cite{Hohenegger:2016yuv} for generic $N$ and $M$) for the K\"ahler parameters (mapping $\omega\to \omega'$) is taken into account. This equality was shown explicitly in \cite{Bastian:2017ing} for $M=1$ and for general $(N,M)$ in \cite{Haghighat:2018gqf} assuming a certain limit of two deformation parameters $\epsilon_{1,2}$ that are required to render $\mathcal{Z}_{N,M}(\omega,\epsilon_{1,2})$ well defined. Combined with the triality of supersymmetric gauge theories argued in \cite{Bastian:2017ary}, this lead to the conjecture \cite{Bastian:2018dfu} of a vast web of dual gauge theories with gauge groups $[U(N')]^{M'}$ with $NM=N'M'$ and $\text{gcd}(N,M)=\text{gcd}(N',M')$.

While the dualities described above link very different looking theories in a non-perturbative fashion, they also imply symmetries for individual theories: focusing on $M=1$, it was first remarked in \cite{Bastian:2018jlf} that this web of dualities implies invariance of the partition function $\mathcal{Z}_{N,M}(\omega)$ under the group $\widetilde{\mathbb{G}}(N)\cong\mathbb{G}(N)\times \mathcal{S}_N$. Here $\mathcal{S}_N\cong \text{Dih}_{N}\subset S_N$ is (a subgroup of) the Weyl group of the largest gauge group that can be engineered in a gauge theory from $X_{N,1}$. Furthermore
\begin{align}
\mathbb{G}(N)\cong\left\{\begin{array}{lcl}\text{Dih}_3 & \text{if} & N=1,3\,, \\ \text{Dih}_2 & \text{if} & N=2\,,\\\text{Dih}_\infty & \text{if} & N\geq 4\,,\end{array}\right.\label{FirstIntroDihedral}
\end{align}
where $\text{Dih}_\infty$ is the group that is freely generated by two elements of order 2.

Using this insight, the properties of the partition function $\mathcal{Z}_{N,1}$ (or more specifically, the free energy $F_{N,1}$ associated with it) have been re-examined in \cite{Paper1,Paper2}: among other things, on the one hand side it was argued in \cite{Paper1} that a particular subsector (called the \emph{reduced} free energy) of $F_{N,1}$ is invariant with respect to the paramodular group $\Sigma_N$, which in the Nekrasov-Shatashvili limit \cite{Nekrasov:2009rc,Mironov:2009uv} $\epsilon_2\to 0$ is extended to $\Sigma_N^*\subset Sp(4,\mathbb{Q})$. This result agrees with the earlier observation in \cite{Ahmed:2017hfr} that the states contributing to the reduced free energy form a symmetric orbifold CFT and that an expansion of the former shows a very characteristic Hecke structure. On the other hand, by examining the examples $N=2,3,4$ up to order $3$ in the instanton parameter $Q_R$ (from the point of view of the largest gauge group $U(N)$ that can be engineered from $X_{N,1}$) it was discussed in \cite{Paper2} that the remaining contributions to the free energy exhibit very suggestive structures. In particular, it was argued, that these BPS counting functions can be written using particular combinations of generating functions $T(\widehat{a}_1,\ldots,\widehat{a}_{N-1};\rho)$ of multiple divisor sums, which were introduced in \cite{Bachmann:2013wba}. These functions are a generalisation of quasi-Jacobi forms and combine in a natural fashion the modular properties (acting on the modular parameter $\rho$) of the free energy with the $\widehat{\mathfrak{a}}_{N+1}$ (affine) gauge algebra (with roots $\widehat{a}_{1,\ldots,N-1}$). 

As was already remarked in \cite{Bachmann:2013wba}, the generating functions $T(\widehat{a}_1,\ldots,\widehat{a}_{N-1};\rho)$ are also intimately connected to multiple zeta values and can be expanded in terms of (reduced) polylogarithms. The latter have in recent years attracted a lot of attention in the study of (loop) amplitudes in (supersymmetric) string theories (see \emph{e.g.} \cite{Schlotterer:2012ny,Broedel:2013aza,Broedel:2013tta,Stieberger:2013wea,Broedel:2014vla,Broedel:2015hia,DHoker:2015wxz,MatthesPhD,Brown1,Brown2,Broedel:2018izr,Zerbini:2018sox} and references therein for an overview). The current paper is motivated by extending this connection and extracting a class of functions from the free energy $F_{N,1}$, which show a certain resemblance of so-called (modular) \emph{graph functions} (see \cite{Broedel:2015hia,DHoker:2015wxz,DHoker:2016mwo,DHoker:2017pvk,Zerbini:2018sox,Zerbini:2018hgs,Gerken:2018jrq,Mafra:2019ddf,Mafra:2019xms,Gerken:2019cxz}) that have been studied recently in the literature. To this end, we consider the so-called unrefined limit $\epsilon_1=-\epsilon_2=\epsilon$ and study instanton expansions of the free energy for $N=2,3$ up to order $Q_R^3$:
\begin{align}
F_{N,1}(\widehat{a}_1,\ldots,\widehat{a}_{N-1},\rho,R,S,\epsilon)=\sum_{r=1}^\infty Q_R^r\sum_{s=0}^\infty \epsilon^{2s-2}\,B^{(N,r)}_{(s)}(\widehat{a}_1,\ldots,\widehat{a}_{N-1},\rho,S)\,.\label{FreeEnergyInstantongExpansionIntro}
\end{align}
The latter exhibit a number of very interesting patterns, which can schematically be represented in the form of graphs, that resemble effective higher point functions (even similar to Feynman diagrams): indeed, to leading instanton order (\emph{i.e.} for $r=1$ in (\ref{FreeEnergyInstantongExpansionIntro})) we find that we can write
\begin{align}
&B^{(N,r=1)}_{(s)}=H_{(s)}^{(0,1)}(\rho,S)\,\sum_{i=0}^{N-1}(W_{(0)}(\rho,S))^{N-1-i}\,(H_{(0)}^{(0,1)}(\rho,S))^i\,\mathcal{O}^{(N),i}(\widehat{a}_{1,\ldots,N-1},\rho)\,,\label{GenFreeEnergyOrd1Intro}
\end{align}
which we have verified for $N=2,3$ up to order $s=4$ and also partially for $N=4$, using the data for $s=0$ provided in \cite{Paper2}. In (\ref{GenFreeEnergyOrd1Intro}) we have as the fundamental building blocks $H_{(s)}^{(0,1)}=B^{(N=1,r=1)}_{(s)}$ and $W_{(0)}$, which is a particular quasi-Jacobi form of weight 0 and index $1$ (see (\ref{Sect:PropOrbN2OrderE0}) for the precise definition). The latter are 'coupled' through the coupling function $\cpf{N}{i}$, which is independent of $S$ and $s$ and depends on the roots of the gauge algebra $\widehat{\mathfrak{a}}_{N+1}$. The form (\ref{GenFreeEnergyOrd1Intro}) can graphically be presented as an $N$-point function, where the external legs are given by $H_{(s)}^{(0,1)}$ and $W_{(0)}$ respectively, which are connected via $\cpf{N}{i}$ (see \figref{Fig:SummaryNPointFct}). While a priori nothing more than an amusing graphical representation, this interpretation seems to go beyond a mere mnemonic device: for the examples $N=2,3,4$ we find that $\cpf{N}{0}=N$, while $\cpf{N}{1}$ is in fact (up to a $\widehat{a}_{1,\ldots,N-1}$-independent term) related to the scalar Greens function on the torus (see appendix~\ref{App:TorusPropagator}). Moreover, the real part of $\cpf{N}{1}$ is the fundamental building block for the study of graph functions in \cite{DHoker:2015wxz}. Higher $\cpf{N}{i}$, for $i>1$, are more involved. However, using the presentation in terms of the $T(\widehat{a}_1,\ldots,\widehat{a}_{N-1};\rho)$ mentioned above, they can be written as combinations of polylogarithms -- a property they share in common with the graph functions in \cite{DHoker:2015wxz}. We leave it to further work to analyse these functions in detail. 

This paper is organised as follows: In section~\ref{Sect:Review} we review the instanton partition function $\mathcal{Z}_{N,1}(\omega,\epsilon_{1,2})$ and free energy $F_{N,1}(\omega,\epsilon_{1,2})$ as well as their symmetries found in \cite{Paper1,Paper2}. In Sections~\ref{Sect:ExN2} and \ref{Sect:ExN3} we study the cases $N=2$ and $N=3$ respectively, which reveal the patterns mentioned above. In Section~\ref{Sect:GeneralStructureSummary} we summarise these examples and formulate a conjecture for the general structure in the form of (\ref{FreeEnergyInstantongExpansionIntro}). We also discuss similarities of the free energy with modular graph functions in the literature of scattering amplitudes in string- and field theory. Section~\ref{Sect:Conclusions} contains our conclusions and an outlook for future work. Furthermore, two appendices contain a short review on modular objects and the generating functions of multiple divisor sums.


\section{Review: Free Energy}\label{Sect:Review}
\subsection{Little String Free Energy}
In order to set the stage for subsequent sections, we start by reviewing the partition function and free energy of a class of little string theories of A-type \cite{Hohenegger:2015btj,Hohenegger:2016eqy,Hohenegger:2016yuv,Bastian:2017ing}. The latter can be engineered through F-theory compactified on a class of toric, non-compact Calabi-Yau threefolds \cite{Kanazawa:2016tnt}, which are called $X_{N,M}$ for $N,M\in\mathbb{N}$. The corresponding non-perturbative BPS partition function $\mathcal{Z}_{N,M}(\omega,\epsilon_{1,2})$ of these LSTs is captured by the topological string partition function on $X_{N,M}$ \cite{Hohenegger:2015btj} (see also \cite{Haghighat:2013gba,Haghighat:2013tka,Hohenegger:2013ala,Hohenegger:2015cba}), which can be computed in a very efficient manner using the refined topological vertex (see \cite{Aganagic:2003db,Hollowood:2003cv,Iqbal:2007ii}): indeed, in \cite{Bastian:2017ing} a general building block 
$W^{\alpha_1,\ldots,\alpha_N}_{\beta_{1},\ldots,\beta_{N}}(\widehat{a}_{1,\ldots,N},S;\epsilon_{1,2})$ was computed that is labelled by $2N$ sets of integer partitions $\alpha_{1,\ldots,N}$ and $\beta_{1,\ldots,N}$ and that depends on (a subset of) the K\"ahler parameters of $X_{N,M}$. The latter allows to compute the full instanton partition function by 'gluing together' $M$-copies of itself, weighted in a suitable manner by the instanton parameters. In the case $M=1$, only a single building block is necessary and the non-perturbative part of the partition function can be written as
\begin{align}
&\mathcal{Z}_{N,1}(\widehat{a}_{1,\ldots,N-1},\rho,S,R;\epsilon_{1,2})=\sum_{\{\alpha\}}\left(\prod_{i=1}^N Q_{m_i}^{|\alpha_i|}\right)\,W^{\alpha_1,\ldots,\alpha_N}_{\alpha_{1},\ldots,\alpha_{N}}(\widehat{a}_{1,\ldots,N-1},\rho,S;\epsilon_{1,2})\,.\label{DefBuildingZ}
\end{align}
Concerning the K\"ahler parameters of $X_{N,1}$, we use the same basis that was introduced in \cite{Bastian:2017ary}, \emph{i.e.} $(\widehat{a}_1,\ldots,\widehat{a}_{N-1},\rho,S,R)$, where the $\widehat{a}_{1,\ldots,N-1}$ play the roles of roots of the affine $\widehat{\mathfrak{a}}_{N+1}$ gauge algebra, $S$ plays the role of a mass parameter, while $R$ is related to the coupling constant. The $Q_{m_i}=e^{2\pi i m_i}$ with $m_i=m_i(R_,S,\widehat{a}_{1,\ldots,N-1},\rho)$ appearing in (\ref{DefBuildingZ}) are linear combinations of $(\widehat{a}_{1,\ldots,N-1},S,R)$ and contain the only dependence on $R$. Thus, (\ref{DefBuildingZ}) is essentially an expansion in powers of $Q_R=e^{2\pi i R}$, which can therefore be identified with an instanton expansion.

Rather than working with the partition function $\mathcal{Z}_{N,1}$, we work with the free energy
\begin{align}
F_{N,1}(\widehat{a}_{1,\ldots,N-1},\rho,S,R;\epsilon_{1,2})=\text{PLog}\,\mathcal{Z}_{N,1}(\widehat{a}_{1,\ldots,N-1},\rho,S,R;\epsilon_{1,2})\,,\label{PlethLog}
\end{align}
which affords the following Fourier expansion 
\begin{align}
F_{N,1}(\widehat{a}_1,\ldots,\widehat{a}_N,S,R;\epsilon_{1,2})=\sum_{s_1,s_2=0}^\infty\sum_{r=0}^\infty\sum_{i_1,\ldots,i_N}^\infty\sum_{k\in\mathbb{Z}}\epsilon_{1}^{s_1-1}\epsilon_{2}^{s_2-1}f^{(s_1,s_2)}_{i_1,\ldots,i_N,k,r}\,Q_{\widehat{a}_1}^{i_1}\ldots Q_{\widehat{a}_N}^{i_N}\,Q_S^k\,Q_R^r\,,\label{TaylorFreeEnergy}
\end{align}
with the coefficients $f^{(s_1,s_2)}_{i_1,\ldots,i_N,k,r}$.\footnote{Due to the plethystic logarithm $\text{PLog}$, $F_{N,1}$ only counts single particle BPS states. The former is defined as $\text{PLog}\,\mathcal{Z}_{N,1}(\widehat{a}_{1,\ldots,N-1},\rho,S,R;\epsilon_{1,2})=\sum_{k=1}^\infty\frac{\mu(k)}{k}\,\ln\mathcal{Z}_{N,1}(k\,\widehat{a}_{1,\ldots,N},k\,S,k\,R;k\,\epsilon_{1,2})$, where $\mu$ is the M\"obius function.} Here we are using the same notation as in \cite{Paper1,Paper2}
\begin{align}
&Q_{\widehat{a}_i}=e^{2\pi i\widehat{a}_i}\,,&&Q_\rho=e^{2\pi i\rho}\,,&&Q_S=e^{2\pi i S}\,,&&Q_R=e^{2\pi i R}\,,&&\text{for} &&i=1,\ldots,N-1\,,
\end{align}
and we introduce the following partial expansion
\begin{align}
H_{(s_1,s_2)}^{(i_1,\ldots,i_N,r)}(\rho,S)&=\sum_{\ell=0}^\infty\sum_{k\in\mathbb{Z}} f^{(s_1,s_2)}_{i_1+\ell,i_2+\ell,\ldots,i_N+\ell,k,r}\,Q_S^k\,Q_\rho^\ell\,,&&\forall i_{1,\ldots,N}\in\mathbb{N}\cup\{0\}\,.\label{DefinitionH}
\end{align}
The free energy can be formally\footnote{Here we do not worry about convergence: we shall reformulate the free energy in (modular) objects, for which we utilise a suitable summation procedure \cite{Weil} to guarantee convergence. See \cite{Paper2} for more details.} recovered as
\begin{align}
F_{N,1}(\widehat{a}_1,\ldots,\widehat{a}_N,S,R;\epsilon_{1,2})=\sum_{s_1,s_2=0}^\infty\sum_{r=0}^\infty \epsilon_1^{s_1-1}\epsilon_2^{s_2-1}\,Q_R^r \,B^{N,r}_{(s_1,s_2)}(\widehat{a}_1,\ldots,\widehat{a}_{N-1},\rho,S)\,,\label{FreeEnergyB}
\end{align}
with the functions
\begin{align}
&B^{(N,r)}_{(s_1,s_2)}=\sum'_{i_1,\ldots,i_N}H_{(s_1,s_2)}^{(i_1,\ldots,i_N,r)}(\rho,S)\,Q_{\widehat{a}_1}^{i_1}\ldots Q_{\widehat{a}_N}^{i_N}\,,&&\text{with} &&Q_\rho=\prod_{i=1}^NQ_{\widehat{a}_i}\,,
\end{align}
and the sum is understood to run over all $(i_1,\ldots,i_N)$ such that (at least) one $i_a=0$. Finally, we define the coefficients which appear in the expansion of $F_{N,1}$ in the unrefined case $\epsilon_1=-\epsilon_2=\epsilon$
\begin{align}
&H_{(s)}^{(i_1,\ldots,i_N,r)}=\sum_{s_1+s_2=2s}(-1)^{s_2-1}\,H_{(s_1,s_2)}^{(i_1,\ldots,i_N,r)}\,,&&B_{(s)}^{(N,r)}=\sum_{s_1+s_2=2s}(-1)^{s_2-1}\,B_{(s_1,s_2)}^{(N,r)}\,.
\end{align}
 
Numerous examples of $H_{(s_1,s_2)}^{(i_1,\ldots,i_N,r)}$ have been computed in \cite{Paper2} and its properties have been discussed in detail. In particular, it has been argued that a suitable set of functions to represent the $H_{(s_1,s_2)}^{(i_1,\ldots,i_N,r)}$ are the generating functions of multiple divisor sums $T(X_1,\ldots,X_\ell;\rho)$ that have been first introduced in \cite{Bachmann:2013wba} and which are defined in (\ref{DefTBachmann}). Rather than recounting all the properties of $B_{(s_1,s_2)}^{(N,r)}$ found in \cite{Paper2}, we shall instead give as an example the simplest case (namely $N=1$) in the following, whose expansion will provide us with the general building blocks that are used in the remainder of this paper.

\subsection{Free Energy for $N=1$}\label{Sect:ExpansionN1Qr}
The unrefined free energy for $N=1$ (with $\epsilon_1=-\epsilon_2=\epsilon$) is of the form
\begin{align}
&F_{1,1}(\rho,S,R;\epsilon_1=-\epsilon_2=\epsilon)=\sum_{s=0}^\infty \epsilon^{2s-2}\sum_{r=0}^\infty  Q_R^r H_{(s)}^{(0,1)}(\rho,S)\,, &&\text{with}&& H_{(s)}^{(0,1)}=\sum_{s_1+s_2=2s}(-1)^{s_2-1}\,H_{(s_1,s_2)}^{(0,1)}\,.\nonumber
\end{align}
To order $Q_R^1$ we have concretely
\begin{align}
\sum_{s=0}^\infty \epsilon^{2s-2} H_{(s)}^{(0,1)}=-\frac{\phi_{-2,1}}{\epsilon^2}+\frac{\phi_{0,1}}{24}-\sum_{n=1}^\infty \epsilon^{2n}\,\frac{(-1)^n\,B_{2n+2}\,E_{2n+2}(\rho) \phi_{-2,1}(\rho,s)  }{(2n-1)!!(2n+2)!!}\,,\label{ExpansionN1FreeEnergy}
\end{align}
while for higher orders in $Q_R^r$ (with $r>1$) it was found in \cite{Ahmed:2017hfr} 
\begin{align}
H_{(s)}^{(0,k)}(\rho,S)=\mathcal{T}_k\left(H^{(0,1)}_{(s)}(\rho,S)\right)=\sum_{a|k}a^{w-1}\,\mu(a)\,\mathcal{H}_{k/a}\left(H^{(0,1)}_{(s)}(a\rho,aS)\right)\,,\label{DefOperatorT}
\end{align}
where $\mathcal{H}_k$ is the $k$-th Hecke operator (see eq.~(\ref{DefHeckeGeneric})) and $\mu$ is the M\"obius function. For later convenience, we explicitly tabulate the first few coefficient-functions up to order $Q_R^2$ as follows
\begin{center}
\begin{tabular}{|c|r|r|}\hline
&&\\[-14pt]
$s$ & $H_{(s)}^{(0,1)}$ & $H_{(s)}^{(0,2)}$ \\[4pt]\hline\hline
&&\\[-14pt]
0 & $-\phi_{-2,1}$ & $-\frac{\phi_{-2,1}}{24}\left(\phi_{0,1}+\psi_2\phi_{-2,1}\right)$ \\[4pt]\hline
&&\\[-14pt]
1 & $\frac{\phi_{0,1}}{24}$ & $\frac{1}{576}\left[24 E_4 \phi_{-2,1}^2+(\phi_{0,1}-2\psi_2\phi_{2,1})(\phi_{0,1}+3\psi_2\phi_{-2,1})\right]$ \\[4pt]\hline
&&\\[-14pt]
2 & $-\frac{E_4\,\phi_{-2,1}}{240}$ & $\frac{\phi_{-2,1}}{11520}\left[5\psi_2^2(\phi_{0,1}+4\psi_2\phi_{-2,1})-2 E_4 (11\phi_{0,1}+41\psi_2\phi_{2,1})\right]$ \\[4pt]\hline
&&\\[-14pt]
3 & $-\frac{E_6\,\phi_{-2,1}}{6048}$ & $\frac{\phi_{-2,1}}{290304}\left[252E_4^2\phi_{-2,1}+87E_4\psi_2\phi_{0,1}-2(8E_6\psi_2\phi_{-2,1}+\psi_2^3(11\phi_{0,1}+7\psi_2\phi_{-2,1}))\right]$ \\[4pt]\hline
&&\\[-14pt]
4 & $-\frac{E_4^2\,\phi_{-2,1}}{172800}$ & \parbox{13cm}{$\frac{\phi_{-2,1}}{74649600}\big[-2178E_4^2\phi_{0,1}+9372E_4 E_6\phi_{-2,1}+\psi_2(240E_6\phi_{0,1}-416 E_6\psi_2\phi_{-2,1}$\\${}\hspace{8.1cm}+105\psi_2^3\phi_{0,1}-242\psi_2^4\phi_{-2,1})\big]$} \\[14pt]\hline
\end{tabular}
\end{center}
The Eisenstein series $E_{2k}$ and $\psi_2$ and the Jacobi forms $\phi_{-2,1}$ and $\phi_{0,1}$ are defined in appendix~\ref{App:ModularStuff}.
\section{Example $N=2$}\label{Sect:ExN2}
We start by considering the free energy for $N=2$. More concretely, using the same notation as in \cite{Paper1,Paper2}, we can separate the latter into two pieces (for $r\geq 1$)
\begin{align}
&B^{(2,r)}_{(s)}=H^{(0,0,r)}_{(s)}(\rho,S)+K^{(r,2)}_{(s)}(\rho,S,\widehat{a}_1)\,,&&\text{with} &&K^{(1,2)}_{(s)}=\sum_{n=1}^\infty H^{(n,0,r)}_{(s)}\left(Q^n_{\widehat{a}_1}+\frac{Q_\rho^n}{Q^n_{\widehat{a}_1}}\right)\,.\label{ExpansionFreeEnergyN2r}
\end{align}
Below we shall analyse the orders $r=1,2,3$ in more detail.  
\subsection{Order $Q_R^1$}
For low values of $s$ we find for the terms $H^{(n,0,1)}_{(s)}$  in (\ref{ExpansionFreeEnergyN2r}) the following expressions\\[-36pt]
\begin{wrapfigure}{l}{\dimexpr 0.35\textwidth + 2\FrameSep + 2\FrameRule\relax}
\vspace{0.5cm}
\begin{shaded*}\raggedleft
\scalebox{1}{\parbox{5.6cm}{\begin{tikzpicture}[scale = 1.50]
\draw (-1,0) circle (0.05);
\draw (1,0) circle (0.05);
\draw[ultra thick] (-0.95,0) -- (0,0) -- (0.95,0);
\draw[thick,fill=gray!60!white] (0,0) circle (0.41);
\node at (-1.45,0) {$H_{(0)}^{(0,1)}$};
\node at (1.45,0) {$H_{(s)}^{(0,1)}$};
\node at (0,0) {$\propNonOrb$};
\end{tikzpicture}}}
\caption{\sl Coupling function $\cpf{2}{1}$ appearing in the $\epsilon$-expansion of $K^{(1,2)}_{(s)}$.}
\label{Fig:PropagatorN2Simp}
\end{shaded*}
\vspace{-3.75cm}
\end{wrapfigure}
\begin{align}
H^{(n,0,1)}_{(0)}&=-\frac{2n}{1-Q_\rho^n}\,\phi_{-2,1}^2\,,\nonumber\\
H^{(n,0,1)}_{(1)}&=\frac{n}{12(1-Q_\rho^n)}\,\phi_{0,1}\,\phi_{-2,1}\,,\nonumber\\
H^{(n,0,1)}_{(2)}&=-\frac{n}{120(1-Q_\rho^n)}\,E_4\,\phi_{-2,1}^2\,,\nonumber\\
H^{(n,0,1)}_{(3)}&=-\frac{n}{3024(1-Q_\rho^n)}\,E_6\,\phi_{-2,1}^2\,,\nonumber\\
H^{(n,0,1)}_{(4)}&=-\frac{n}{86400(1-Q_\rho^n)}\,E_4^2\,\phi_{-2,1}^2\,.\nonumber
\end{align}

\noindent
Comparing with the $H_{(s)}^{(0,1)}$ appearing in the expansion of the free energy of $N=1$ (as tabulated in section~\ref{Sect:ExpansionN1Qr}), these examples exhibit a pattern which suggests\footnote{Our notation is explained in more detail in section~\ref{Sect:GeneralStructureSummary}: there we shall denote general coupling functions $\cpf{N}{i}$, where $i$ is a summation index labelling different classes of couplings.}
\begin{align}
&K^{(1,2)}_{(s)}=H_{(0)}^{(0,1)}\,\propNonOrb\,H_{(s)}^{(0,1)}\,,&&\text{where} &&\propNonOrb(\widehat{a}_1,\rho)=-\sum_{n=1}^\infty\frac{2n}{1-Q_\rho^n}\,\left(Q^n_{\widehat{a}_1}+\frac{Q_\rho^n}{Q^n_{\widehat{a}_1}}\right)\,.\label{Sect:PropN2OrderE0}
\end{align}
This relation can be represented graphically as in \figref{Fig:PropagatorN2Simp}, where $H_{(0)}^{(0,1)}$ is coupled to $H_{(s)}^{(0,1)}$ through the coupling function $\propNonOrb$. Notice that the latter is independent of $s$ and only depends on $(\widehat{a}_1,\rho)$, but not $S$ (whose dependence in \figref{Fig:PropagatorN2Simp} is only given through $H_{(s)}^{(0,1)}$ and $H_{(0)}^{(0,1)}$).

Similar to $K^{(1,2)}_{(s)}$, we can compute $H^{(0,0,1)}_{(s)}$, where the first few examples are\\[-30pt]
\begin{wrapfigure}{l}{\dimexpr 0.35\textwidth + 2\FrameSep + 2\FrameRule\relax}
\vspace{0.25cm}
\begin{shaded*}\raggedleft
\scalebox{1}{\parbox{5.6cm}{\begin{tikzpicture}[scale = 1.50]
\draw (-1,0) circle (0.05);
\draw (1,0) circle (0.05);
\draw[ultra thick] (-0.95,0) -- (0,0) -- (0.95,0);
\draw[ultra thick,fill=black] (0,0) circle (0.05);
\node at (-1.4,0) {$W_{(0)}$};
\node at (1.45,0) {$H_{(s)}^{(0,1)}$};
\draw[thick,fill=gray!60!white] (0,0) circle (0.41);
\node at (0,0) {$\cpf{2}{0}$};
\end{tikzpicture}}}
\caption{\sl Coupling function $\propOrb$ appearing in the expansion of the free energy $H^{(1,2)}_{(s)}$ in powers of $\epsilon$. }
\label{Fig:PropagatorN2Simpn0}
%
\end{shaded*}
\vspace{-4cm}
\end{wrapfigure}
\begin{align}
H^{(0,0,1)}_{(0)}&=-\frac{1}{12}\,\phi_{-2,1}\,(\phi_{0,1}+2E_2\,\phi_{-2,1})\,,\nonumber\\
H^{(0,0,1)}_{(1)}&=\frac{1}{288}\,\phi_{0,1}\,(\phi_{0,1}+2E_2\,\phi_{-2,1})\,,\nonumber\\
H^{(0,0,1)}_{(2)}&=-\frac{1}{2880}\,E_4\,\phi_{-2,1}\,(\phi_{0,1}+2E_2\,\phi_{-2,1})\,,\nonumber\\
H^{(0,0,1)}_{(3)}&=-\frac{1}{72576}\,E_6\,\phi_{-2,1}\,(\phi_{0,1}+2E_2\,\phi_{-2,1})\,,\nonumber\\
H^{(0,0,1)}_{(4)}&=-\frac{1}{2073600}\,E_4^2\,\phi_{-2,1}\,(\phi_{0,1}+2E_2\,\phi_{-2,1})\,.\nonumber
\end{align}

\noindent
In the same manner as (\ref{Sect:PropN2OrderE0}), these examples suggest the pattern
\begin{align}
&H^{(0,0,1)}_{(s)}(\rho,S)=W_{(0)}\,\propOrb\,H_{(s)}^{(0,1)}\,,&&\text{where} &&\left\{\begin{array}{l}\propOrb =2\,, \\ W_{(0)}=\frac{1}{24}\,(\phi_{0,1}+2E_2\,\phi_{-2,1})\,.\end{array}\right.\label{Sect:PropOrbN2OrderE0}
\end{align}
As above, this relation can be represented graphically as in \figref{Fig:PropagatorN2Simpn0}, where $W_{(0)}$ is coupled to $H_{(s)}^{(0,1)}$ through $\propOrb$. The quasi-Jacobi form $W_{(0)}$ is the leading term in the expansion
\begin{align}
W(\rho,S,\epsilon_{1,2})=\frac{\theta_1(\rho;S+\epsilon_-)\theta_1(\rho;S-\epsilon_-)-\theta_1(\rho;S+\epsilon_+)\theta_1(\rho;S-\epsilon_+)}{\theta_1(\rho;\epsilon_1)\theta_1(\rho;\epsilon_2)}=W_{(0)}+\mathcal{O}(\epsilon_1,\epsilon_2)\,,\nonumber
\end{align}
which appeared in \cite{Hohenegger:2015btj} (see also \cite{Paper1}), as the function governing the counting of BPS states of a (local) M5-brane with a single M2-brane ending on it on either side. We refer the reader to \cite{Hohenegger:2015btj,Paper1} for more details.

In (\ref{Sect:PropN2OrderE0}) and (\ref{Sect:PropOrbN2OrderE0}) we have called $\propNonOrb$ and $\propOrb$ coupling functions and have graphically represented the contribution to the free energy in \figref{Fig:PropagatorN2Simp} and \figref{Fig:PropagatorN2Simpn0} like a(n effective) 'propagator' appearing in a two-point function. In the case of the former, this connection can be made more concrete: $\propNonOrb$ can (up to a non-holomorphic contribution) be related to (the derivative of) the scalar two-point function on the torus, as we shall explain explicitly in Section~\ref{Sect:PropN2}.

\subsection{Order $Q_R^2$}
We have seen that to leading instanton order, the free energy can be presented as a 'two-point function' involving the effective coupling functions $\propOrb$ and $\propNonOrb$. Higher orders in $Q_R$ are more involved, however, still exhibit very interesting similar structures. 

Indeed, slightly adopting the notation used in \cite{Paper2}, for $r=2$ in (\ref{ExpansionFreeEnergyN2r}), we can write
\begin{align}
&H^{(n,0,2)}_{(s)}=\sum_{i=1}^4 \left[g^{i,(n,2)}_{(s)}+\theta_{n,2}\,h^{i,(n,2)}_{(s)}\right]\,\frac{n\,\phi_{-2,1}^{5-i} \phi_{0,1}^{i-1}}{1-Q_\rho^n}\,,&&\text{with} &&\theta_{a,b}=\left\{\begin{array}{lcl}0 & \text{if} & \text{gcd}(a,b)=1\,,\\ 1 & \text{if} & \text{gcd}(a,b)>1\,,\end{array}\right.\nonumber
\end{align}
where we can tabulate the coefficients $g^{i,(n,2)}_{(s)}$ as follows
\begin{center}
\begin{tabular}{|c|r|r|r|r|}\hline
&&&&\\[-14pt]
$s$ & $g^{1,(n,2)}_{(s)}$ & $g^{2,(n,2)}_{(s)}$ & $g^{3,(n,2)}_{(s)}$ & $g^{4,(n,2)}_{(s)}$  \\[4pt]\hline\hline
&&&&\\[-14pt]
0 & $-\frac{n^4+2E_4}{24}$ & $-\frac{n^2}{12}$ & $-\frac{1}{96}$ & $0$ \\[4pt]\hline
&&&&\\[-14pt]
1 & $\frac{8n^6+182n^2E_4-85 E_6}{5040}$ & $\frac{41 E_4+34 n^4}{2880}$ & $\frac{5n^2}{576}$ & $\frac{1}{2304}$ \\[4pt]\hline
&&&&\\[-14pt]
2 & $\frac{2300E_6 n^2-1722E_4^2-1386E_4 n^4-11 n^8}{362880}$ & $\frac{-595E_4 n^2+290E_6-52 n^6}{120960}$ & $\frac{-34E_4-59 n^4}{69120}$ & $\frac{-n^2}{4608}$  \\[4pt]\hline
&&&&\\[-14pt]
3 & \parbox{4cm}{$\frac{-5 E_6 (3539 E_4 + 2376 n^4)+7 n^{10}}{19958400}$\\${}$\hspace{0.95cm}$+\frac{E_4 n^2 (889 E_4 + 82 n^4)}{604800}$} & $\frac{42 E_4 (10 E_4 + 9 n^4)-628 E_6 n^2+7 n^8}{870912}$ & $\frac{217 E_4 n^2-105 E_6+43 n^6}{1451520}$ & $\frac{n^4}{55296}$  \\[14pt]\hline
\end{tabular}
\end{center}

\noindent 
and in a similar fashion, we can tabulate the coefficients $h^{i,(n,2)}_{(s)}$ as follows\hspace{20cm}$\phantom{x}$\\[2pt]
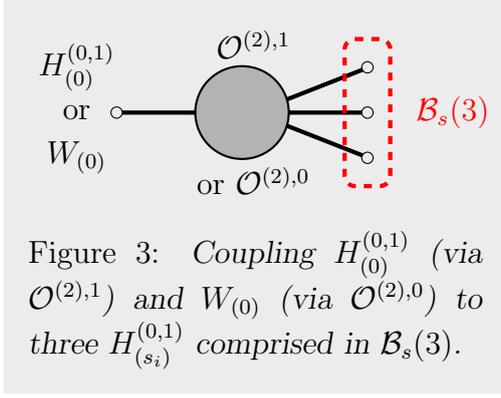
\begin{wrapfigure}{l}{\dimexpr 0.35\textwidth + 2\FrameSep + 2\FrameRule\relax}
\vspace{-3.1cm}
\begin{shaded*}
\centering
\scalebox{1}{\parbox{6cm}{\begin{tikzpicture}[scale = 1.50]
\draw (-1.2,0) circle (0.05);
\draw[ultra thick] (-1.15,0) -- (0.95,0);
\draw[ultra thick,fill=black] (-0.1,0) circle (0.05);
\node at (-1.55,0.4) {$H_{(0)}^{(0,1)}$};
\node at (-1.55,0) {\text{or}};
\node at (-1.55,-0.4) {$W_{(0)}$};
\draw[ultra thick] (0,0) -- (0.96,0.38);
\draw[ultra thick] (0,0) -- (0.96,-0.38);
\draw (1,0.4) circle (0.05);
\draw (1,0) circle (0.05);
\draw (1,-0.4) circle (0.05);
\node[red] at (1.75,0) {$\mathcal{B}_s(3)$};
\draw[thick,fill=gray!60!white] (-0.1,0) circle (0.41);
\draw[ultra thick,rounded corners,dashed,red] (0.8,0.65)--(1.2,0.65)--(1.2,-0.65) -- (0.8,-0.65) -- cycle;
\node at (0,0.6) {$\cpf{2}{1}$};
\node at (0,-0.6) {\text{or} $\cpf{2}{0}$};
\end{tikzpicture}}}
\caption{\sl Coupling $H^{(0,1)}_{(0)}$ (via $\cpf{2}{1}$) and $W_{(0)}$ (via $\cpf{2}{0}$) to three $H^{(0,1)}_{(s_i)}$ comprised in $\mathcal{B}_s(3)$.}
\label{Fig:PropagatorN2Background3}
\end{shaded*}
\vspace{-2.1cm}
\end{wrapfigure}
\hspace{0.2cm}
\begin{tabular}{|c|r|r|r|r|}\hline
&&&&\\[-14pt]
$s$ & $h^{1,(n,2)}_{(s)}$ & $h^{2,(n,2)}_{(s)}$ & $h^{3,(n,2)}_{(s)}$ & $h^{4,(n,2)}_{(s)}$ \\[4pt]\hline\hline
&&&&\\[-14pt]
0 & $\frac{\psi_2^2}{72}$ & $-\frac{\psi_2}{72}$ & $\frac{1}{288}$ & $0$ \\[4pt]\hline
&&&&\\[-14pt]
1 & $\frac{E_6+\psi_2^3}{432}$ & $\frac{3E_4-4\psi_2^2}{1728}$ & $\frac{\psi_2}{1728}$ & $-\frac{1}{6912}$ \\[4pt]\hline
&&&&\\[-14pt]
2 & $\frac{8E_6 \psi_2+11 \psi_2^4}{51840}$ & $\frac{-8E_6-11 \psi_2^3}{51840}$ & $\frac{5 \psi_2^2-4E_4}{69120}$ & $0$ \\[4pt]\hline
&&&&\\[-14pt]
3 & $\frac{\psi_2^3(11 \psi_2^2-12 E_4)}{435456}$ & $\frac{\psi_2^2(12 E_4-11 \psi_2^2)}{435456}$ & $\frac{8 E_6+7 \psi_2^3}{1741824}$ & $0$ \\[4pt]\hline
\end{tabular}


${}$\\[-10pt]

\noindent
We can similarly treat the contribution $H^{(0,0,2)}_{(s)}$ in (\ref{ExpansionFreeEnergyN2r}). Adapting the notation of \cite{Paper2}, we have 
\begin{align}
H^{(0,0,2)}_{(s)}(\rho,S)=\sum_{i=1}^5 u^{i,(2,2)}_{(s)}(\rho)\,\phi_{-2,1}^{5-i}(\rho,S)\,\phi_{0,1}^{i-1}(\rho,S)\,,
\end{align}
where we find for the leading terms in $s$\\[-30pt]
\begin{center}
\begin{tabular}{|c||r|r|r|r|r|}\hline
&&&&&\\[-14pt]
$s$ & $u^{1,(2,2)}_{(s)}$ & $u^{2,(2,2)}_{(s)}$ & $u^{3,(2,2)}_{(s)}$ & $u^{4,(2,2)}_{(s)}$ & $u^{5,(2,2)}_{(s)}$  \\[4pt]\hline\hline
&&&&&\\[-14pt]
0 & $\frac{2 E_2  \left(\psi_2 ^2-6 E_4 \right)+10 E_6 -\psi_2 ^3}{1728}$ & $\frac{\psi_2  (\psi_2 -2 E_2 )-3 E_4 }{1728}$ & $\frac{-4 E_2 -\psi_2 }{6912}$ & $-\frac{1}{6912}$ & $0$\\[4pt]\hline
&&&&&\\[-14pt]
1 & $\frac{4E_2  \left(7 \psi_2 ^3-44 E_6 \right)+3 E_4 (68 E_4 + 35 \psi_2^2) -42 \psi_2^4}{145152}$ & $\frac{E_2(276  E_4 -40  \psi_2 ^2)-146 E_6 +5 \psi_2 ^3}{207360}$ & $\frac{\psi_2  (4 E_2 -5 \psi_2 )+18 E_4 }{82944}$ & $\frac{4 E_2 +\psi_2 }{288\cdot 24^2}$ & $\frac{1}{165888}$\\[4pt]\hline
%
\end{tabular}
\end{center}
Due to their complexity, we refrain from writing higher terms explicitly.

Combining these explicit expressions suggests that $B^{2,2}_{(s)}$ can be presented as
{\allowdisplaybreaks
\begin{align}
&B^{(2,2)}_{(s)}(\rho,S,\widehat{a}_1)=\mathcal{T}_2\left(H^{(0,0,1)}_{(s)}\right)+2\,\propNonOrb\mathcal{H}_2\left[H_{(0)}^{(0,1)}\,H^{(0,1)}_{(s)}\right]-2^{2s-4}\,K^{(1,2)}_{(s)}(2\rho,2S,2\widehat{a}_1)\nonumber\\
&\hspace{2cm}+\left[H^{(0,1)}_{(0)}\,\cpf{2}{1}-\frac{1}{4}\,W_{(0)}\,\cpf{2}{0}\right]\,\mathcal{B}_s(3)+H^{(0,1)}_{(0)}\sum_{\underline{s}=\{s_1,s_2,s_3\}} \cpf{2}{1}_{2,\underline{s}}\, \prod_{i=1}^3H^{(0,1)}_{(s_i)}\,,\label{N2OrdR2NonOrb}
\end{align}}
where we used the shorthand notation $\underline{s}=\{s_1,s_2,s_3\}$. The terms in the first line of (\ref{N2OrdR2NonOrb}) are obtained through particular operations from the contributions (\ref{Sect:PropN2OrderE0}) or (\ref{Sect:PropOrbN2OrderE0}) (\emph{i.e.} the free energy to order $Q_R^{1}$). The first term in the second line corresponds to a coupling of $H_{(0)}^{(0,0)}$ (through $\cpf{2}{1}$) or $W_{(0)}$ (through $\cpf{2}{0}$) to three $H_{(s_i)}^{(0,1)}$ forming $\mathcal{B}_s(3)$, as shown in \figref{Fig:PropagatorN2Background3}. The latter is built from three $H^{(0,1)}_{(s_i)}$ in the following fashion
\begin{align}
\mathcal{B}_s(3)=\frac{1}{8}\sum_{{s_1,s_2,s_3}\atop s_1+s_2+s_3=s+2} c_{\underline{s}}\,H^{(0,1)}_{(s_1)}\,H^{(0,1)}_{(s_2)}\,H^{(0,1)}_{(s_3)}\,,&&\text{with} && \begin{array}{l}c_{\underline{s}}\in\mathbb{Z}\\\underline{s}=\{s_1,s_2,s_3\}\end{array}\,.\label{DefN2R2background}
\end{align}
Explicitly, for the lowest values of $s$, we find
{\allowdisplaybreaks
\begin{align}
&\mathcal{B}_0(3)=0\,,\hspace{1cm}\mathcal{B}_1(3)=4\,H^{(0,1)}_{(0)} H^{(0,1)}_{(1)} H^{(0,1)}_{(2)} - 3\,(H^{(0,1)}_{(0)})^2 H^{(0,1)}_{(3)}\,,\nonumber\\
&\mathcal{B}_2(3)=4\, (H^{(0,1)}_{(1)})^2 H^{(0,1)}_{(2)} + 6 H^{(0,1)}_{(0)} (H^{(0,1)}_{(2)})^2 + 12 H^{(0,1)}_{(0)} H^{(0,1)}_{(1)} H^{(0,1)}_{(3)} + 2 (H^{(0,1)}_{(0)})^2 H^{(0,1)}_{(4)}\,,\nonumber\\
&\mathcal{B}_3(3)=15 (H^{(0,1)}_{(1)})^2 H^{(0,1)}_{(3)} + H^{(0,1)}_{(0)} H^{(0,1)}_{(2)} H^{(0,1)}_{(3)} + 100 H^{(0,1)}_{(0)} H^{(0,1)}_{(1)} H^{(0,1)}_{(4)} + 192 (H^{(0,1)}_{(0)})^2 H^{(0,1)}_{(5)}\,.\nonumber
\end{align}}

\noindent
${}$\\[-36pt]
The explicit numerical coefficients appearing in $\mathcal{B}_s(3)$ for higher values in $s$ are difficult to predict (and are in fact not unique), thus constituting information encoded in the free energy, that does not immediately follow from the case $N=1$. Finally, the last line in (\ref{N2OrdR2NonOrb}) represents new terms that involve similar sums as in (\ref{Sect:PropN2OrderE0}), with higher powers in $n$, as well as Eisenstein series. It is, however, amusing that this contribution can still be represented in a diagrammatic  

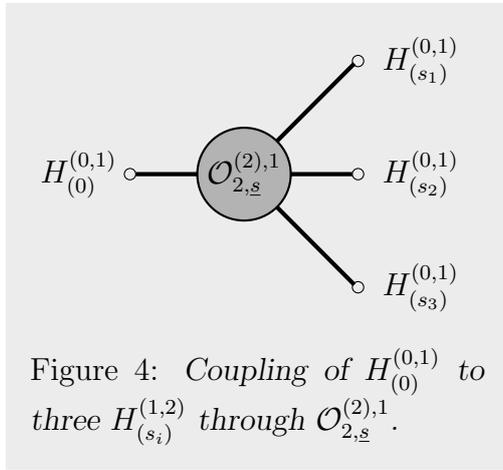
\begin{wrapfigure}{l}{\dimexpr 0.35\textwidth + 2\FrameSep + 2\FrameRule\relax}
\vspace{-0.8cm}
\begin{shaded*}
\scalebox{1}{\parbox{5.8cm}{\begin{tikzpicture}[scale = 1.50]
\draw (-1,0) circle (0.05);
\draw (1,1) circle (0.05);
\draw (1,0) circle (0.05);
\draw (1,-1) circle (0.05);
\draw[ultra thick] (-0.95,0) -- (0,0) -- (0.95,0);
\draw[ultra thick] (0,0) -- (0.9675,0.9775);
\draw[ultra thick] (0,0) -- (0.9675,-0.9775);
\draw[ultra thick,fill=black] (0,0) circle (0.05);
\node at (-1.45,0) {$H_{(0)}^{(0,1)}$};
\node at (1.55,1) {$H_{(s_1)}^{(0,1)}$};
\node at (1.55,0) {$H_{(s_2)}^{(0,1)}$};
\node at (1.55,-1) {$H_{(s_3)}^{(0,1)}$};
\draw[thick,fill=gray!60!white] (0,0) circle (0.41);
\node at (0,0) {$\cpf{2}{1}_{2,\underline{s}}$};
\end{tikzpicture}}}
\caption{\sl Coupling of $H_{(0)}^{(0,1)}$ to three $H^{(1,2)}_{(s_i)}$ through $\cpf{2}{1}_{2,\underline{s}}$.}
\label{Fig:VertexN2states3}
%
\end{shaded*}
\vspace{-1cm}
\end{wrapfigure}

\noindent 
fashion along the lines of the remaining terms: as shown in \figref{Fig:VertexN2states3}, it can be represented as $H_{(0)}^{(0,0)}$ coupling to  three other  $H^{(0,1)}_{(s_i)}$ through the coupling function $\cpf{2}{1}_{2,\underline{s}}$.\footnote{Our notation is explained in section~\ref{Sect:GeneralStructureSummary}: we denote $\cpf{N}{i}_{r,\underline{s}}$, where $i$ is a summation index labelling different classes of couplings, $r$ denotes the order of $Q_R^r$, while $\underline{s}=\{s_1,\ldots,s_k\}$ is a set of labels governing the $\epsilon$-expansion.} For given $s$, the latter is
\begin{align}
\cpf{2}{1}_{2,\underline{s}}=\sum_{n=1}^\infty\frac{d_{s;\underline{s}} n^\kappa}{1-Q_\rho^n}\,\left(Q^n_{\widehat{a}_1}+\frac{Q_\rho^n}{Q^n_{\widehat{a}_1}}\right)+\mathfrak{w}^{r=2}_{s,\underline{s}}(\rho)\,.\label{CouplingN2r2cpf}
\end{align}
In the first term, we have $s=\frac{\kappa-5}{2}+s_1+s_2+s_3$ and the $d_{s;\underline{s}}\in\mathbb{Q}$ are numerical coefficients. The second term in (\ref{CouplingN2r2cpf}) is independent of $\widehat{a}_1$ and $\mathfrak{w}^{r=2}_{s,\underline{s}}(\rho)$ are modular forms of weight $w$, which satisfy $s_1+s_2+s_3+\frac{w}{2}=s+3$. To leading order in $s$, we find explicitly \\[18pt]

\vspace{-1cm}
\begin{center}
\parbox{5cm}{\begin{tabular}{|c||c||r|r|}\hline
&&&\\[-16pt]
$s$ & $\underline{s}$ & $d_{s;\underline{s}}$ & $\mathfrak{w}^2_{s,\underline{s}}$  \\[2pt]\hline\hline
&&&\\[-16pt]
0 & $\{0, 0, 1\}$ & $2$ & $0$\\[2pt]\hline
&&&\\[-16pt]
 & $\{0 ,0 , 0 \}$ & $\frac{-1}{24}$ & $0$\\[2pt]\hline\hline
 &&&\\[-16pt]
1 & $\{0, 1, 1\}$ & $5$ & $\frac{-4 E_4}{15}$\\[2pt]\hline
 &&&\\[-16pt]
 & $\{ 0 , 0 , 2\}$ & $\frac{26}{3}$ & $\frac{-10 E_4}{63}$\\[2pt]\hline
&&&\\[-16pt]
& $\{ 0 , 0 , 1\}$ & $\frac{-17}{60}$ & $\frac{19 E_6}{630}$\\[2pt]\hline
 &&&\\[-16pt]
 & $\{0 , 0 , 0\}$ & $\frac{1}{630}$ & $0$\\[2pt]\hline\hline
&&&\\[-16pt]
2 & $\{1 , 1 , 1\}$ & $3$ & $\frac{-4E_4}{15}$\\[2pt]\hline
&&&\\[-16pt]
 & $\{0 , 1 , 2\}$ & $\frac{85}{3}$ & $0$\\[2pt]\hline
\end{tabular}}
\hspace{0.35cm}
\parbox{5.4cm}{\begin{tabular}{|c||c||r|r|}\hline
&&&\\[-16pt]
$s$ & $\underline{s}$ & $d_{s;\underline{s}}$ & $\mathfrak{w}^2_{s,\underline{s}}$ \\[2pt]\hline\hline
&&&\\[-16pt]
2 & $\{0 , 0 , 3\}$ & $\frac{115}{3}$ & $\frac{14 E_4}{15}$\\[2pt]\hline
 &&&\\[-16pt]
& $\{0 , 1 , 1\}$ & $\frac{-59}{120}$ & $\frac{-E_6}{105}$\\[2pt]\hline
 &&&\\[-16pt]
 & $\{0 , 0 , 2\}$ & $\frac{-11}{12}$ & $0$\\[2pt]\hline
 &&&\\[-16pt]
 & $\{0 , 0 , 1\}$ & $\frac{13}{1260}$ & $\frac{E_4^2}{1260}$\\[2pt]\hline
 &&&\\[-16pt]
 & $\{0 , 0 , 0\}$ & $\frac{-11}{362880}$ & $0$\\[2pt]\hline\hline
 &&&\\[-16pt]
3 & $\{1 , 1 , 2\}$ & $\frac{62}{3}$ & $\frac{-10 E_4}{7}$\\[2pt]\hline
 &&&\\[-16pt]
 & $\{0 , 1 , 3\}$ & $\frac{314}{3}$ & $0$\\[2pt]\hline
  &&&\\[-16pt]
 & $\{0 , 0 , 4\}$  & $254$ & $0$\\[2pt]\hline
\end{tabular}}
\hspace{0.35cm}
\parbox{5.4cm}{\vspace{-0.8cm}\begin{tabular}{|c||c||r|r|}\hline
&&&\\[-16pt]
$s$ & $\underline{s}$ & $d_{s;\underline{s}}$ & $\mathfrak{w}^2_{s,\underline{s}}$ \\[2pt]\hline\hline
&&&\\[-16pt]
3 & $\{1 , 1 , 1\}$  & $\frac{-1}{4}$ & $\frac{-5E_6}{126}$\\[2pt]\hline
 &&&\\[-16pt]
 & $\{0 , 1 , 2\}$  & $\frac{-5}{2}$ & $\frac{61E_6}{1386}$\\[2pt]\hline
&&&\\[-16pt]
 & $\{0 , 0 , 3\}$  & $\frac{-18}{5}$ & $\frac{91E_6}{990}$\\[2pt]\hline
&&&\\[-16pt]
 & $\{0 , 1 , 1\}$  & $\frac{43}{2520}$ & $0$\\[2pt]\hline
 &&&\\[-16pt]
 & $\{0 , 0 , 2\}$  & $\frac{41}{1260}$ & $\frac{5 E_4^2}{616}$\\[2pt]\hline
 &&&\\[-16pt]
 & $\{0 , 0 , 1\}$  & $\frac{-1}{5184}$ & $0$\\[2pt]\hline
 &&&\\[-16pt]
 & $\{0 , 0 , 0\}$  & $\frac{1}{2851200}$ & $0$\\[2pt]\hline
\end{tabular}}
\end{center}
We stress that the $d_{s;\underline{s}}$ and $\mathfrak{w}^2_{s,\underline{s}}$ are in general not unique\footnote{This is \emph{e.g.} due to linear relations such as \emph{e.g.} $E_4 H_{(0)}^{(0,1)}=240 H_{(2)}^{(0,1)}$.} and the above table simply gives a(n economic) presentation. We finally remark that, despite being structurally very similar, we have chosen to present the terms involving $\cpf{2}{0,1}$ and $\cpf{2}{1}_{2,\underline{s}}$ separately in the last line of (\ref{N2OrdR2NonOrb}) since the latter is a completely holomorphic function, while the former (once the $\cpf{2}{1}$ are summed up) also contain quasi-modular contributions (see eq.~(\ref{N2SimpPropRev}) below). From a 'Feynman diagrammatic' point of view, however, these two represent two similar classes of couplings.

\subsection{Order $Q_R^3$}
We can continue the previous analysis to order $Q_R^3$. However, since explicit expansions to this order are very difficult to compute, we shall limit ourselves to only the leading orders in $s$. Due to their complexity, we refrain from explicitly writing the $H_{(s)}^{(n,0,3)}(\rho,S)$, but we refer the reader to \cite{Paper2}. The leading two orders in $s$, however, suggest the following presentation
{\allowdisplaybreaks
\begin{align}
&B^{(2,3)}_{(s)}(\rho,S,\widehat{a}_1)=\mathcal{T}_3\left(H^{(0,0,1)}_{(s)}\right)+3\,\propNonOrb\,\mathcal{H}_3\left[H_{(0)}^{(0,1)}\,H_{(s)}^{(0,1)}\right]-3^{2s-4}\,K_{(s)}^{(1,2)}(3\rho,3S,3\widehat{a}_1)\nonumber\\
&\hspace{1cm}+\left[H_{(0)}^{(0,1)}\,\cpf{2}{1}\,-\frac{1}{9}W_{(0)}\,\cpf{2}{0}\right]\,\mathcal{B}_s(5)+H^{(0,1)}_{(0)}\sum_{\underline{s}=\{s_1,s_2,s_3,s_4,s_5\}} \cpf{2}{1}_{3,\underline{s}}\, \prod_{i=1}^5H^{(0,1)}_{(s_i)}\,,\label{N2OrdR3NonOrb}
\end{align}}
which directly generalises (\ref{N2OrdR2NonOrb}) to order $Q_R^3$: the terms in the first line are obtained through certain operators from the free energy at order $Q_R^1$. Furthermore, the first term in the second line of (\ref{N2OrdR3NonOrb}) couples $H_{(0)}^{(0,1)}$ (through $\cpf{2}{1}$) and $W_{(0)}$ (through $\cpf{2}{0}$) to $\mathcal{B}_s(5)$, as schematically shown in \figref{Fig:PropagatorN2Background5}. To leading  order in $s$, we find for the former:

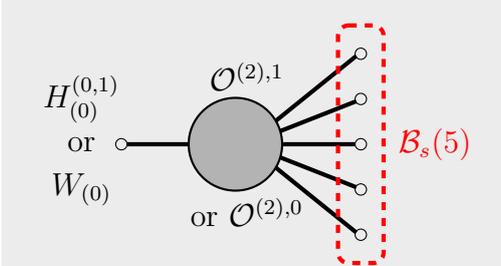
\begin{wrapfigure}{l}{\dimexpr 0.35\textwidth + 2\FrameSep + 2\FrameRule\relax}
\vspace{-0.8cm}
\begin{shaded*}
\centering
\scalebox{1}{\parbox{5.8cm}{\begin{tikzpicture}[scale = 1.50]
\draw (-1.1,0) circle (0.05);
\draw[ultra thick] (-1.05,0) -- (0.95,0);
\draw[ultra thick,fill=black] (-0.2,0) circle (0.05);
\node at (-1.45,0.4) {$H_{(0)}^{(0,1)}$};
\node at (-1.45,0) {\text{or}};
\node at (-1.45,-0.4) {$W_{(0)}$};
\draw (1,0.8) circle (0.05);
\draw (1,0.4) circle (0.05);
\draw (1,0) circle (0.05);
\draw (1,-0.4) circle (0.05);
\draw (1,-0.8) circle (0.05);
\node[red] at (1.65,0) {$\mathcal{B}_s(5)$};
\draw[ultra thick] (0,0) -- (0.96,0.78);
\draw[ultra thick] (0,0) -- (0.96,0.38);
\draw[ultra thick] (0,0) -- (0.96,-0.38);
\draw[ultra thick] (0,0) -- (0.96,-0.78);
\draw[thick,fill=gray!60!white] (-0.1,0) circle (0.41);
\draw[ultra thick,rounded corners,dashed,red] (0.8,1.05)--(1.2,1.05)--(1.2,-1.05) -- (0.8,-1.05) -- cycle;
\node at (0,0.6) {$\cpf{2}{1}$};
\node at (0,-0.6) {\text{or} $\cpf{2}{0}$};
\end{tikzpicture}}}
\caption{\sl Coupling $H^{(0,1)}_{(0,0)}$ (via $\cpf{2}{1}$) and $W_{(0)}$ (via $\cpf{2}{0}$) to five $H^{(0,1)}_{(s_i)}$ comprised in $\mathcal{B}_s(5)$.}
\label{Fig:PropagatorN2Background5}
\end{shaded*}
\vspace{-1.4cm}
\end{wrapfigure}

${}$\\[-50pt]
\begin{align}
&\mathcal{B}_0(5)=0\,,\nonumber\\
&\mathcal{B}_1(5)=\frac{64}{11}H_{(0)}^{(0,1)}\big[22 (H_{(1)}^{(0,1)})^3 H_{(2)}^{(0,1)}\nonumber\\
&\hspace{0.3cm}+99 H_{(0)}^{(0,1)}(H_{(1)}^{(0,1)})^2 H_{(3)}^{(0,1)}+103 (H_{(0)}^{(0,1)})^2 H_{(2)}^{(0,1)} H_{(3)}^{(0,1)}\nonumber\\
&\hspace{0.3cm}+330 (H_{(0)}^{(0,1)})^2 H_{(1)}^{(0,1)} H_{(4)}^{(0,1)}\big]\,.\label{LeadingBackground5}
\end{align}
Finally, the last term in the second line in (\ref{N2OrdR3NonOrb}) generalises the contribution depicted in \figref{Fig:VertexN2states3} in the sense that it couples $H^{(0,1)}_{(0)}$ to five different $H^{(0,1)}_{(s_i)}$ (rather than

\begin{wrapfigure}{r}{\dimexpr 0.35\textwidth + 2\FrameSep + 2\FrameRule\relax}
\vspace{-0.9cm}
\begin{shaded*}
\scalebox{1}{\parbox{5.8cm}{\begin{tikzpicture}[scale = 1.50]
\draw (-1,0) circle (0.05);
\draw (1,1.2) circle (0.05);
\draw (1,0.6) circle (0.05);
\draw (1,0) circle (0.05);
\draw (1,-0.6) circle (0.05);
\draw (1,-1.2) circle (0.05);
\draw[ultra thick] (-0.95,0) -- (0,0) -- (0.95,0);
\draw[ultra thick] (0,0) -- (0.97,1.17);
\draw[ultra thick] (0,0) -- (0.965,0.585);
\draw[ultra thick] (0,0) -- (0.965,-0.585);
\draw[ultra thick] (0,0) -- (0.97,-1.17);
\draw[ultra thick,fill=black] (0,0) circle (0.05);
\node at (-1.45,0) {$H_{(0)}^{(0,1)}$};
\node at (1.55,1.2) {$H_{(s_1)}^{(0,1)}$};
\node at (1.55,0.6) {$H_{(s_2)}^{(0,1)}$};
\node at (1.55,0) {$H_{(s_3)}^{(0,1)}$};
\node at (1.55,-0.6) {$H_{(s_4)}^{(0,1)}$};
\node at (1.55,-1.2) {$H_{(s_5)}^{(0,1)}$};
\draw[thick,fill=gray!60!white] (0,0) circle (0.41);
\node at (0,0) {$\cpf{2}{1}_{3,\underline{s}}$};
\end{tikzpicture}}}
\caption{\sl Coupling of $H_{(0)}^{(0,1)}$ to five $H^{(1,2)}_{(s_i)}$ through $\cpf{2}{1}_{3,\underline{s}}$.}
\label{Fig:VertexN2states5}
%
\end{shaded*}
\vspace{-1.5cm}
\end{wrapfigure}
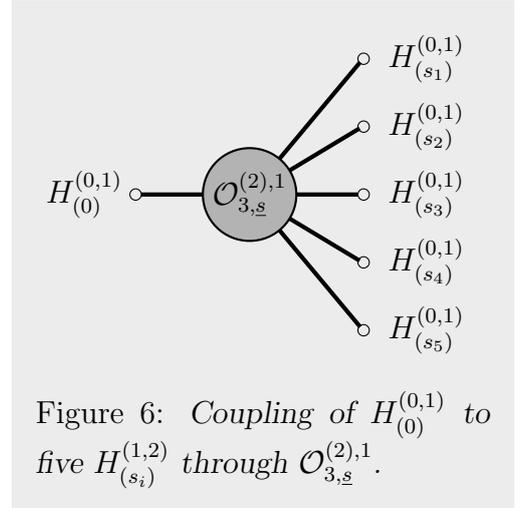

\noindent
 just 3) through a coupling function $\cpf{2}{1}_{2,\underline{s}}$, where $\underline{s}=\{s_1,s_2,s_3,s_4,s_5\}$, as schematically shown in \figref{Fig:VertexN2states5}. We can write
\begin{align}
\cpf{2}{1}_{3,\underline{s}}=\sum_{n=1}^\infty\frac{d_{s;\underline{s}} n^\kappa}{1-Q_\rho^n}\,\left(Q^n_{\widehat{a}_1}+\frac{Q_\rho^n}{Q^n_{\widehat{a}_1}}\right)+\mathfrak{w}^{r=2}_{s,\underline{s}}(\rho)\,,\label{N2r3cpf}
\end{align}
where $d_{s,\underline{s}}\in \mathbb{Q}$ are numerical coefficient and we have $s=\frac{\kappa-9}{2}+\sum_{a=1}^5s_a$. Furthermore, the $\mathfrak{w}^{r=2}_{s,\underline{s}}(\rho)$ are independent of $\widehat{a}_1$ and are modular forms of weight $w$, which satisfy $\sum_{a=1}^5 s_a+\frac{w}{2}=s+5$. The leading contributions in $s$ can be tabulated in the following:\\[-10pt]

\begin{center}
\parbox{6.5cm}{\begin{tabular}{|c||c||r|r|}\hline
&&&\\[-16pt]
$s$ & $\underline{s}$ & $d_{s,\underline{s}}$ & $\mathfrak{w}^{r=2}_{s,\underline{s}}$  \\[2pt]\hline\hline
&&&\\[-16pt]
0 & $\{0, 0 , 1 , 1 , 1\}$ & $-\frac{32}{2}$ & $0$\\[2pt]\hline
&&&\\[-16pt]
 &$\{ 0 , 0 , 0 , 1 , 2\}$ & $-64$ & $0$\\[2pt]\hline
&&&\\[-16pt]
 & $\{ 0 , 0 , 0 , 0 , 3\}$ & $-\frac{176}{3}$ & $0$\\[2pt]\hline
&&&\\[-16pt]
 & $\{0 , 0 , 0 , 1 , 1\}$ & $\frac{4}{3}$ & $0$\\[2pt]\hline
&&&\\[-16pt]
 & $\{0 , 0 , 0 , 0 , 2\}$ & $\frac{14}{9}$ & $0$\\[2pt]\hline
&&&\\[-16pt]
 & $\{ 0 , 0 , 0 , 0 ,  1\}$ & $-\frac{4}{135}$ & $0$\\[2pt]\hline
&&&\\[-16pt]
 & $\{ 0 , 0 , 0 , 0 , 0\}$ & $\frac{1}{7560}$ & $0$\\[2pt]\hline\hline
&&&\\[-16pt]
1 & $\{ 0 , 1 , 1 , 1 , 1\}$ & $-\frac{80}{3}$ & $-\frac{16}{5}\,E_4$\\[2pt]\hline
&&&\\[-16pt]
 & $\{0 , 0 , 1 , 1 , 2\}$ & $-\frac{1264}{3}$ & $-\frac{32}{21}\,E_4$\\[2pt]\hline
\end{tabular}}
\hspace{1.5cm}
\parbox{7cm}{\begin{tabular}{|c||c||r|r|}\hline
&&&\\[-16pt]
$s$ & $\underline{s}$ & $d_{s,\underline{s}}$ & $\mathfrak{w}^{r=2}_{s,\underline{s}}$  \\[2pt]\hline\hline
 &&&\\[-16pt]
 & $\{0 , 0 , 0 , 2 , 2\}$ & $-\frac{5504}{9}$ & $\frac{2000}{693}\,E_4$\\[2pt]\hline
&&&\\[-16pt]
 & $\{0 , 0 , 0 , 1 , 3\}$ & $-\frac{2960}{3}$ & $\frac{1024}{165}\,E_4$\\[2pt]\hline
 &&&\\[-16pt]
 & $\{ 0 , 0 , 1 , 1 , 1\}$ & $\frac{104}{15}$ & $-\frac{32}{315}\,E_6$\\[2pt]\hline
&&&\\[-16pt]
 & $\{ 0 , 0 , 0 , 1 , 2\}$ & $30$ & $0$\\[2pt]\hline
  &&&\\[-16pt]
 & $\{ 0 , 0 , 0 , 0 , 3\}$ & $\frac{908}{45}$ & $\frac{56}{495}\,E_6$\\[2pt]\hline
&&&\\[-16pt]
 & $\{0 , 0 , 0 , 1 , 1\}$ & $-\frac{34}{105}$ & $0$\\[2pt]\hline
   &&&\\[-16pt]
 & $\{0 , 0 , 0 , 0 , 2\}$ & $-\frac{284}{945}$ & $0$\\[2pt]\hline
 &&&\\[-16pt]
 & $\{0 , 0 , 0 , 0 , 1\}$ & $\frac{1}{280}$ & $0$\\[2pt]\hline
  &&&\\[-16pt]
 & $\{ 0 , 0 , 0 , 0 , 0\}$ & $-\frac{4}{467775}$ & $0$\\[2pt]\hline
\end{tabular}}

\end{center}

\noindent
As in the case of $\cpf{2}{1}_{2,\underline{s}}$, however, we stress that the above coefficients are not unique, but just represent an economic choice. 

\subsection{Nekrasov-Shatashvili Limit}
The discussion above was exclusively focused on the so-called unrefined limit, \emph{i.e.} $\epsilon_{1}=-\epsilon_2$. For completeness, we also briefly comment on another limit, namely the Nekrasov-Shatashvili limit \cite{Nekrasov:2009rc,Mironov:2009uv}, \emph{i.e.} effectively $\epsilon_2\to 0$. However, we shall limit ourselves to studying the order $Q_R^1$ (up to order $\epsilon_1^5$): while the latter also exhibits some very interesting patterns, extracting a coupling function akin to (\ref{Sect:PropN2OrderE0}) (in the unrefined limit) is more involved. We shall therefore leave an in-depth analysis of this limit to future work~\cite{SHIqbal}.

We start with a presentation of the free energy (\ref{TaylorFreeEnergy}) like in (\ref{ExpansionFreeEnergyN2r}), suitable for the NS-limit
\begin{align}
&B^{(2,r)}_{(s,0)}=H^{(0,0,r)}_{(s,0)}(\rho,S)+K^{(r,2)}_{(s,0)}(\rho,S,\widehat{a}_1)\,,&&\text{with} &&K^{(1,2)}_{(s,0)}=\sum_{n=1}^\infty H^{(n,0,r)}_{(s,0)}(\rho,S)\left(Q^n_{\widehat{a}_1}+\frac{Q_\rho^n}{Q^n_{\widehat{a}_1}}\right)\,.\nonumber
\end{align}
The contributions $H^{(0,0,r)}_{(s,0)}$ (see (\ref{DefinitionH})) have been studied in \cite{Ahmed:2017hfr} and have been shown to exhibit a very particular Hecke symmetry. We therefore focus on the contribution $K^{(r,2)}_{(s,0)}(\rho,S,\widehat{a}_1)$, or more precisely $H^{(n,0,r)}_{(s,0)}$. In order to reveal similar structures as in the unrefined limit (see eq.~(\ref{Sect:PropN2OrderE0})) we first need to compute the $H_{(s,0)}^{(0,1)}$. The latter can be written as \cite{Paper2}
\begin{align}
H_{(s,0)}^{(0,1)}=u_{(s,0)}^{1,(1,1)}(\rho)\,\phi_{-2,1}(\rho,S)+u_{(s,0)}^{2,(1,1)}(\rho)\,\phi_{0,1}(\rho,S)\,,
\end{align}
where $u_{(s,0)}^{i,(1,1)}(\rho)$ are quasi-modular forms of weight $s$ (for $i=1$) and $s-2$ (for $i=2$). For low values of $s$, they can be tabulated as follows
\begin{center}
\begin{tabular}{|c||r|r|r|r|}\hline
&&&&\\[-16pt]
$s$ & $0$ & $2$ & $4$ & $6$  \\[2pt]\hline\hline
&&&&\\[-16pt]
$u_{(s,0)}^{1,(1,1)}$ & $-1$ & $-\frac{E_2}{48}$  & $-\frac{5 E_2^2 + 13 E_4}{40\cdot 24^2}$ & $-\frac{184 E_6+273 E_2 E_4 +35 E_2^3 }{70\cdot 24^4}$\\[2pt]\hline
&&&&\\[-16pt]
$u_{(s,0)}^{2,(1,1)}$ & $0$ & $\frac{1}{96}$  & $\frac{E_2}{8\cdot 24^2}$  & $\frac{7 E_4+5 E_2^2 }{160\cdot 24^3}$\\[2pt]\hline
\end{tabular}
\end{center}
Notice that the $H_{(s,0)}^{(0,1)}$ satisfy
\begin{align}
\frac{\partial H_{(s,0)}^{(0,1)}}{\partial E_2}=\frac{1}{2\cdot 24}\,H_{(s-1,0)}^{(0,1)}\,.
\end{align}
Furthermore, the $H^{(n,0,r=1)}_{(s,0)}$ (up to $s=6$) have been computed in \cite{Paper2} and are of the form
\begin{align}
H^{(n,0,2)}_{(s,0)}=\sum_{i=1}^2 g^{i,(n,1)}_{(s,0)}\,\frac{\phi_{-2,1}^{3-i} \phi_{0,1}^{i-1}}{1-Q_\rho^n}\,,&&\text{with} &&g^{i,(n,1)}_{(s,0)}=\sum_{\kappa\in\mathbb{N}_{\text{odd}}} p_{\kappa,s}^i(\rho)\,n^\kappa\,,
\end{align}
where $p_{\kappa,s}^i(\rho)$ are quasi-modular forms of weight $s-\kappa-2i+3$, which can be tabulated as follows
\begin{center}
\begin{tabular}{|c||c||c|c||c|c|c||c|c|c|c|c|}\hline
$s$ & 0 & \multicolumn{2}{c||}{2} & \multicolumn{3}{c||}{4} & \multicolumn{4}{c|}{6}  \\[2pt]\hline
$\kappa$ & $1$ & $3$ & $1$ & $5$ & $3$ & $1$ & $7$ & $5$ & $3$ & $1$ \\[2pt]\hline\hline
&&&&&&&&&&\\[-14pt]
$p_{\kappa,s}^1$ & $-2$ & $\frac{1}{3}$ & $\frac{-E_2}{12}$ & $\frac{-1}{60}$ & $\frac{E_2}{72}$ & $\frac{-13 E_4-10 E_2^2}{5760}$ & $\frac{192}{35\cdot 24^3}$ & $\frac{-48 E_2}{5\cdot 24^3}$ & $\frac{2(10E_2^2+13 E_4)}{5\cdot 24^3}$ & $\frac{-92 E_6-273 E_4 E_2-70 E_2^3}{210\cdot 24^3}$\\[2pt]\hline
&&&&&&&&&&\\[-14pt]
$p_{\kappa,s}^2$ & $0$ & $0$ & $\frac{1}{24}$ & $0$ & $\frac{-1}{144}$ & $\frac{E_2}{576}$ & $0$ & $\frac{1}{5\cdot 24^2}$ & $\frac{-E_2}{6\cdot 24^2}$ & $\frac{5(E_4+E_2^2)}{2\cdot 24^3}$ \\[2pt]\hline
&&&&&&&&&&\\[-14pt]
$p_{\kappa,s}^3$ & $0$ & $0$ & $0$ & $0$ & $0$ & $\frac{-1}{8\cdot 24^2}$ & $0$ & $0$ & $\frac{1}{2\cdot 24^3}$ & $\frac{-E_2}{8\cdot 24^3}$ \\[2pt]\hline
\end{tabular}
\end{center}
Using this explicit form, we find that $K^{(1,2)}_{(s,0)}$ can also be written as
\begin{align}
&K^{(1,2)}_{(s,0)}=\sum_{a,b=0}^{s/2} H_{(a,0)}^{(0,1)}\,\mathcal{M}_{ab}(\widehat{a}_1)\,H_{(b,0)}^{(0,1)}\,,&&\text{with}&&\mathcal{M}_{ab}=-2\sum_{n=1}^\infty \frac{(-1)^{\frac{s}{2}+a+b} n^{s+1-2(a+b)}\left(Q^n_{\widehat{a}_1}+\frac{Q_\rho^n}{Q^n_{\widehat{a}_1}}\right)}{(1-Q_\rho^n)\Gamma(s-2(a+b-1))}\,.\nonumber
\end{align}
Here, using a slightly imprecise notation, we understand that $1/\Gamma(-m)=0\hspace{0.2cm}\forall m\in\mathbb{N}\cup \{0\}$. In this way, $\mathcal{M}$ is a $\left(\tfrac{s}{2}+1\right)\times \left(\tfrac{s}{2}+1\right)$ symmetric matrix with zero entries for $a+b>\tfrac{s}{2}$. Furthermore, the entries on the off-diagonal of $\mathcal{M}$ correspond precisely to the coupling function $\propNonOrb$ in (\ref{Sect:PropN2OrderE0}), while all entries in the top-left half correspond to holomorphic derivatives of the latter with respect to $\widehat{a}_1$. While this is still a very intriguing pattern (which shall be further analysed in \cite{SHIqbal}) it does not make the coupling function appear in  such a clean fashion as in the unrefined case (see~eq.(\ref{Sect:PropN2OrderE0})). Therefore, in the following we shall further analyse the latter limit.
\section{Example $N=3$}\label{Sect:ExN3}
Following \cite{Paper1,Paper2}, the free energy for $N=3$ can be decomposed as in (\ref{ExpansionFreeEnergyN2r}) for the case $N=2$
\begin{align}
&B^{(3,r)}_{(s)}(\rho,S,\widehat{a}_1,\widehat{a}_2)=H^{(0,0,0,r)}_{(s)}(\rho,S)+K^{(r,3)}_{(s)}(\rho,S,\widehat{a}_1,\widehat{a}_2)\,,\label{BN3r}
\end{align}
where in the following we shall limit ourselves to $r=1$. The contribution $K^{(r,3)}_{(s)}$ comprises three different types of terms. Using a slightly different notation than in \cite{Paper2} and generalising a pattern arising up to order $s=2$, we can present it in the following form
\begin{align}
&K^{(1,3)}_{(s)}(\rho,S,\widehat{a}_1,\widehat{a}_2)=\sum_{n=1}^\infty \left[n\,\frac{\lambda^1_{s}(\rho,S)}{1-Q_\rho^n}\left(\Delta_+^{(n)}+Q_\rho^n\,\Delta_+^{(-n)}\right)+n^2\,\frac{\lambda^2_{s}(\rho,S)}{(1-Q_\rho^n)^2}\left(\Delta_+^{(n)}+\Delta_+^{(-n)}\right)\right]\nonumber\\
&\hspace{0.1cm}+\frac{\lambda^3_{s}(\rho,S)}{24}\sum_{n_1,n_2=1}^\infty\left[\frac{n_2(n_2+2n_1)}{(1-Q_\rho^{n_1})(1-Q_\rho^{n_2})}+\frac{n_1^2-n_2^2}{(1-Q_\rho^{n_1})(1-Q_\rho^{n_1+n_2})}\right]\sum_{i=1}^3 Q_{\widehat{a}_i}^{n_1+n_2}\sum_{j\neq i}Q_{\widehat{a}_j}^{n_1}\,.\label{N3StructExpanGen}
\end{align} 
In the second line we have defined $Q_{\widehat{a}_3}=Q_\rho/(Q_{\widehat{a}_1}Q_{\widehat{a}_2})$ to keep the notation compact, while $\Delta_+^{(n)}=\sum_{i=1}^3Q_{\widehat{a}_i}^{n}$. Furthermore, the $\lambda^{1,2,3}_s$ are quasi-Jacobi forms of index $3$ and weights\footnote{We remark that the $\lambda^{i}_s$ are independent of the summation variables $n$ or $n_{1,2}$ in (\ref{N3StructExpanGen}).} $w_1=2s-4$, $w_2=w_3=2s-6$, respectively. The $\lambda^i_s$ can be decomposed as
\begin{align}
\lambda_s^i(\rho,S)=-\sum_{a=1}^3\alpha_{a,s}^i(\rho)\,(\phi_{-2,1}(\rho,S))^{4-a}\,(\phi_{0,1}(\rho,S))^{a-1}\,,
\end{align}
where the $\alpha^i_{a,s}$ are quasi-modular forms of weight $w_i+6$ (for $a=1$), $w_i+4$ (for $a=2$) and $w_i+2$ (for $a=3$) respectively. Up to order $s=3$, they can be tabulated as follows
\begin{center}
\begin{tabular}{|c||c|c|c||c|c|c||c|c|c||c|c|c||c|c|c|}\hline
$s$ &  \multicolumn{3}{c||}{0} & \multicolumn{3}{c||}{1} & \multicolumn{3}{c||}{2} & \multicolumn{3}{c||}{3} & \multicolumn{3}{c|}{4}  \\[2pt]\hline
$a$  & $1$ & $2$ & $3$ & $1$ & $2$ & $3$ & $1$ & $2$ & $3$ & $1$ & $2$ & $3$ & $1$ & $2$ & $3$  \\[2pt]\hline\hline
&&&&&&&&&&&&&&&\\[-14pt]
$\alpha_{a,s}^1$   & $\frac{E_2}{6}$ & $\frac{1}{12}$ & $0$ & $0$ & $-\frac{E_2}{144}$ & $-\frac{1}{288}$ & $\frac{E_2 E_4}{1440}$ & $\frac{E_4}{2880}$ & $0$ & $\frac{E_2 E_6}{36288}$ & $\frac{E_6}{72576}$ & $0$ & $\frac{E_2 E_4^2}{1036800}$ & $\frac{E_4^2}{2073600}$ & $0$\\[2pt]\hline
&&&&&&&&&&&&&&&\\[-14pt]
$\alpha_{a,s}^2$    & $1$ & $0$ & $0$ & $0$ & $-\frac{1}{24}$ & $0$ & $\frac{E_4}{240}$ & $0$ & $0$ & $\frac{E_6}{6048}$ & $0$ & $0$ & $\frac{E_4^2}{172800}$ & $0$ & $0$\\[2pt]\hline
&&&&&&&&&&&&&&&\\[-14pt]
$\alpha_{a,s}^3$  & $24$ & $0$ & $0$ & $0$ & $-1$ & $0$ & $\frac{E_4}{10}$ & $0$ & $0$ & $\frac{E_6}{252}$ & $0$ & $0$ & $\frac{E_4^2}{7200}$ & $0$ & $0$\\[2pt]\hline
\end{tabular}
\end{center}

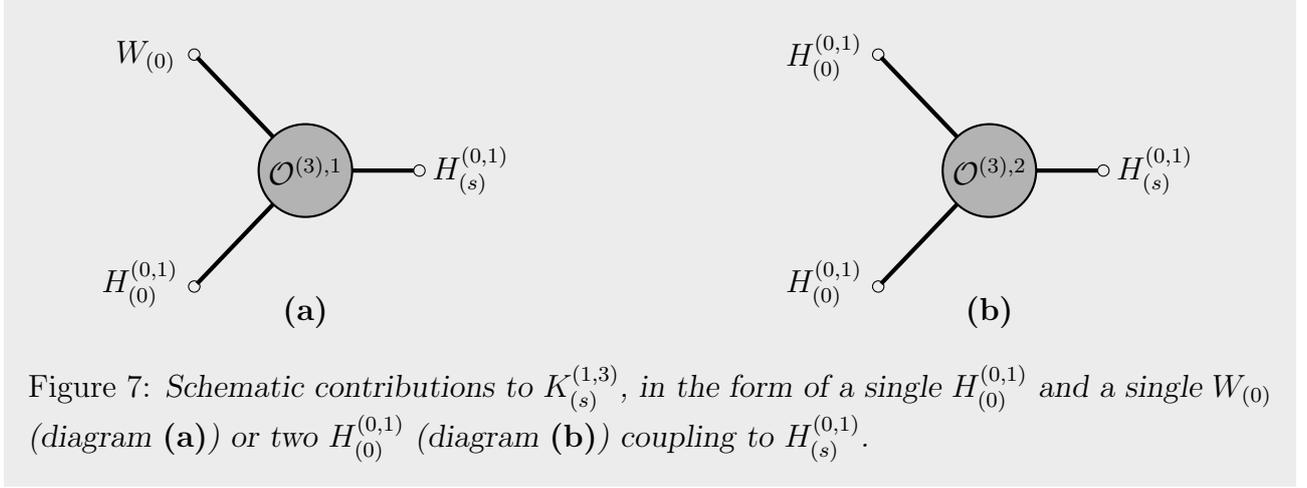
\begin{figure}[h]
\begin{shaded*}
\centering
\scalebox{1}{\parbox{14.7cm}{\begin{tikzpicture}[scale = 1.50]
\draw (-0.975,1.025) circle (0.05);
\draw (-0.975,-1.025) circle (0.05);
\draw (1,0) circle (0.05);
\draw[ultra thick] (-0.95,1) -- (0,0);
\draw[ultra thick] (-0.95,-1) -- (0,0);
\draw[ultra thick] (0,0) -- (0.95,0);
\draw[ultra thick,fill=black] (0,0) circle (0.05);
\node at (-1.4,1) {$W_{(0)}$};
\node at (-1.45,-1) {$H_{(0)}^{(0,1)}$};
\node at (1.45,0) {$H_{(s)}^{(0,1)}$};
\draw[thick,fill=gray!60!white] (0,0) circle (0.41);
\node at (0,0) {$\cpf{3}{1}$};
\node at (0,-1.25) {$\text{\bf(a)}$};
\begin{scope}[xshift=6cm]
\draw (-0.975,1.025) circle (0.05);
\draw (-0.975,-1.025) circle (0.05);
\draw (1,0) circle (0.05);
\draw[ultra thick] (-0.95,1) -- (0,0);
\draw[ultra thick] (-0.95,-1) -- (0,0);
\draw[ultra thick] (0,0) -- (0.95,0);
\draw[ultra thick,fill=black] (0,0) circle (0.05);
\node at (-1.45,1) {$H_{(0)}^{(0,1)}$};
\node at (-1.45,-1) {$H_{(0)}^{(0,1)}$};
\node at (1.45,0) {$H_{(s)}^{(0,1)}$};
\draw[thick,fill=gray!60!white] (0,0) circle (0.41);
\node at (0,0) {$\cpf{3}{2}$};
\node at (0,-1.25) {$\text{\bf(b)}$};
\end{scope}
\end{tikzpicture}}}
\caption{\sl Schematic contributions to $K^{(1,3)}_{(s)}$, in the form of a single $H_{(0)}^{(0,1)}$ and a single $W_{(0)}$ (diagram {\bf (a)}) or two $H_{(0)}^{(0,1)}$ (diagram {\bf (b)}) coupling to $H_{(s)}^{(0,1)}$.}
\label{Fig:ThreePointN3sim}
%
\end{shaded*}
\end{figure}

\noindent
These contributions can be summarised by the following pattern
\begin{align}
K^{(1,3)}_{(s)}=\cpf{3}{1} W_{(0)}\,H_{(0)}^{(0,1)}\,H_{(s)}^{(0,1)}+\cpf{3}{2} H_{(0)}^{(0,1)}\,H_{(0)}^{(0,1)}\,H_{(s)}^{(0,1)}\,,
\end{align}
where we defined
{\allowdisplaybreaks
\begin{align}
\cpf{3}{1}&=-\sum_{n=1}^\infty\frac{2n}{1-Q_\rho^n}\left[\Delta_+^{(n)}+Q_\rho^n\Delta_+^{(-n)}\right]\,,\label{N3StructureCouplingNonOrba}\\
\cpf{3}{2}&=\sum_{n_{1,2}=1}^\infty\frac{\sum_{i} Q_{\widehat{a}_i}^{n_1+n_2}\sum_{j\neq i}Q_{\widehat{a}_j}^{n_1}}{1-Q_\rho^{n_1}}\left[\frac{n_2(n_2+2n_1)}{(1-Q_\rho^{n_2})}+\frac{n_1^2-n_2^2}{(1-Q_\rho^{n_1+n_2})}\right]+\sum_{n=1}^\infty\frac{n^2 Q_\rho^n\,(\Delta_+^{(n)}+\Delta_+^{(-n)})}{(1-Q_\rho^{n})^2}\,\,,\nonumber
\end{align}}
which can be presented as either a $H_{(0)}^{(0,1)}$ and a $W_{(0)}$ or two $H_{(0)}^{(0,1)}$ coupling to $H_{(s)}^{(0,1)}$ (\figref{Fig:ThreePointN3sim}).

\noindent
It remains to study $H^{(1,3)}_{(s)}$ appearing in (\ref{BN3r}), for which the first few examples (in $s$) can be written as follows\\[-36pt]
\begin{wrapfigure}{l}{\dimexpr 0.37\textwidth + 2\FrameSep + 2\FrameRule\relax}
\vspace{0.1cm}
\begin{shaded*}\raggedleft
\scalebox{1}{\parbox{5.5cm}{\begin{tikzpicture}[scale = 1.50]
\draw (-0.975,1.025) circle (0.05);
\draw (-0.975,-1.025) circle (0.05);
\draw (1,0) circle (0.05);
\draw[ultra thick] (-0.95,1) -- (0,0);
\draw[ultra thick] (-0.95,-1) -- (0,0);
\draw[ultra thick] (0,0) -- (0.95,0);
\draw[ultra thick,fill=black] (0,0) circle (0.05);
\node at (-1.4,1) {$W_{(0)}$};
\node at (-1.4,-1) {$W_{(0)}$};
\node at (1.45,0) {$H_{(s)}^{(0,1)}$};
\draw[thick,fill=gray!60!white] (0,0) circle (0.41);
\node at (0,0) {$\cpf{3}{0}$};
\end{tikzpicture}}}
\caption{\sl Contribution to $H^{(0,0,0,1)}_{(s)}$ as two $W_{(0)}$ coupling to $H_{(s)}^{(0,1)}$.}
\label{Fig:ThreePointN3Orb}
%
\end{shaded*}
\vspace{-6cm}
\end{wrapfigure}
\begin{align}
&\nonumber\\[-4pt]
H^{(0,0,0,1)}_{(0)}&=-\frac{1}{192}\, \phi_{-2,1} (\phi_{0,1}+2E_2\,\phi_{-2,1})^2\,,\nonumber\\
H^{(0,0,0,1)}_{(1)}&=\frac{1}{4608}\,\phi_{0,1}\,(\phi_{0,1}+2E_2\,\phi_{-2,1})^2\,,\nonumber\\
H^{(0,0,0,1)}_{(2)}&=-\frac{1}{46080}\,E_4\,\phi_{-2,1}\,(\phi_{0,1}+2E_2\,\phi_{-2,1})^2\,,\nonumber\\
H^{(0,0,0,1)}_{(3)}&=-\frac{1}{1161216}\,E_6\,\phi_{-2,1}\,(\phi_{0,1}+2E_2\,\phi_{-2,1})^2\,,\nonumber\\
H^{(0,0,0,1)}_{(4)}&=-\frac{1}{33177600}\,E_4^2\,\phi_{-2,1}\,(\phi_{0,1}+2E_2\,\phi_{-2,1})^2\,,\nonumber\\
&\nonumber
\end{align}

\noindent
These explicit examples exhibit a clear pattern which suggest that for generic $s$ we can write
\begin{align}
&H^{(0,0,0,1)}_{(s)}=\cpf{3}{0}\,W_{(0)}\,W_{(0)}\,H_{(s)}^{(0,1)}\,,&&\text{where} &&\mathcal{O}^{(3),0}=3\,.
\end{align}
This form can graphically be represented as in \figref{Fig:ThreePointN3Orb}, with two $W_{(0)}$ coupling through the constant $\cpf{3}{0}=3$ to $H_{(s)}^{(0,1)}$.
\section{General Structure of the Unrefined Free Energy}\label{Sect:GeneralStructureSummary}
\subsection{Summary of Examples and General Conjecture}
Studying explicit expansions of the free energy for the cases $N=2$ and $N=3$ in the unrefined limit has revealed a number of very intriguing patterns, which we conjecture to hold in general: we have seen that the coefficient functions $H_{(s)}^{(0,1)}$, which appear in the expansion of the free energy for $N=1$ at order $Q_R^1$ (see eq.~(\ref{ExpansionN1FreeEnergy})) serve as the fundamental building blocks to construct the free energies for $N>1$ (also to higher orders in $Q_R$). In particular we have seen that the way in which these building blocks are combined, has a certain resemblance of a (Feynman) diagrammatic expansion. More specifically, the examples studied above suggest that to leading order in $Q_R$, the free energy (\ref{FreeEnergyB}) in the unrefined limit can be written as
\begin{align}
&B^{(N,r)}_{(s)}=H_{(s)}^{(0,1)}(\rho,S)\,\sum_{i=0}^{N-1}(W_{(0)}(\rho,S))^{N-1-i}\,(H_{(0)}^{(0,1)}(\rho,S))^i\,\mathcal{O}^{(N),i}(\widehat{a}_{1,\ldots,N-1},\rho)\,,\label{GenFreeEnergyOrd1}
\end{align}

\begin{wrapfigure}{l}{\dimexpr 0.45\textwidth + 2\FrameSep + 2\FrameRule\relax}
\vspace{-0.9cm}
\begin{shaded*}\raggedleft
\scalebox{1}{\parbox{7.4cm}{\begin{tikzpicture}[scale = 1.50]
\draw (-0.975,1.025) circle (0.05);
\draw (-0.975,-1.025) circle (0.05);
\draw (1,0) circle (0.05);
\draw[ultra thick] (-0.95,1) -- (0,0);
\draw[ultra thick] (-0.95,-1) -- (0,0);
\draw[ultra thick] (-1.15,-0.75) -- (0,0);
\draw (-1.18,-0.77) circle (0.05);
\draw (-1.385,-0.255) circle (0.05);
\draw (-1.18,0.77) circle (0.05);
\draw (-1.385,0.255) circle (0.05);
\node[rotate=65] at (-1.2,0.5) {$\cdots$};
\draw[ultra thick] (-1.35,-0.25) -- (0,0);
\draw[ultra thick] (-1.35,0.25) -- (0,0);
\node[rotate=-65] at (-1.2,-0.5) {$\cdots$};
\draw[ultra thick] (-1.15,0.75) -- (0,0);
\draw[ultra thick] (0,0) -- (0.95,0);
\draw[ultra thick,fill=black] (0,0) circle (0.05);
\node at (-1.2,1.3) {$W_{(0)}$};
\node at (-1.5,0.95) {$W_{(0)}$};
\node at (-1.75,0.325) {$W_{(0)}$};
\node[rotate=58] at (-2.4,1.2) {\parbox{2cm}{\small $(N-1-i)$ \\[-6pt]${}$\hspace{0.5cm}times}};
\node at (-1.2,-1.3) {$H_{(0)}^{(0,1)}$};
\node at (-1.6,-0.85) {$H_{(0)}^{(0,1)}$};
\node at (-1.8,-0.31) {$H_{(0)}^{(0,1)}$};
\node[rotate=-55] at (-2.2,-1.2) {\parbox{1.3cm}{\small $i$ times}};
\node at (1.45,0) {$H_{(s)}^{(0,1)}$};
\draw[ultra thick,<->] (-1.5,-1.6) to [out=143,in=280] (-2.3,-0.35);
\draw[ultra thick,<->] (-1.5,1.6) to [out=-143,in=-280] (-2.3,0.35);
\draw[thick,fill=gray!60!white] (0,0) circle (0.41);
\node at (0,0) {$\cpf{N}{i}$};
\end{tikzpicture}}}
\caption{\sl Schematic contribution to $B^{(N,1)}_{(s)}$ as $(N-1-i)$  $W_{(0)}$ and $i$ $H_{(0)}^{(0,1)}$ coupling to $H_{(s)}^{(0,1)}$ through $\mathcal{O}^{(N),i}$, for $i=0,\ldots,N-1$.}
\label{Fig:SummaryNPointFct}
%
\end{shaded*}
\vspace{-1cm}
\end{wrapfigure}

\noindent
where each term in this sum can be represented graphically through a diagram of the type shown in \figref{Fig:SummaryNPointFct}: they correspond to the 'coupling' of $i$ factors of $H_{(0)}^{(0,1)}$ and $(N-1-i)$ factors of $W_{(0)}$ to $H_{(s)}^{(0,1)}$ through $\mathcal{O}^{(N),i}(\widehat{a}_{1,\ldots,N-1},\rho,S)$. In this decomposition, the only $s$-dependence appears through $H_{(s)}^{(0,1)}$ (on the right hand side of \figref{Fig:SummaryNPointFct}), while the only dependence on the roots $\widehat{a}_{1,\ldots,N-1}$ of the gauge algebra $\mathfrak{a}_{N+1}$ is located in $\mathcal{O}^{(N),i}$ for $i>0$. Indeed, the summand $i=0$ captures the contribution called $H_{(s)}^{(1,N)}$ in (\ref{ExpansionFreeEnergyN2r}) for $N=2$ and (\ref{BN3r}) for $N=3$, which is independent of the $\widehat{a}_{1,\ldots,N-1}$. In general, extrapolating the examples\footnote{In \cite{Paper2} further results have been published for the case $N=4$, which are also compatible with this form.} encountered in Sections~\ref{Sect:ExN2} and \ref{Sect:ExN3}, the $\mathcal{O}^{(N),i}$ can schematically be written in the following form
\begin{align}
\mathcal{O}^{(N),i}(\widehat{a}_{1,\ldots,N-1},\rho,S)=\sum_\ell\sum_{n_1,\ldots,n_{i}=1}^\infty\frac{p^i_\ell(n_1,\ldots,n_i)\,\Lambda_\ell^i(\widehat{a}_1,\ldots,\widehat{a}_{N-1},\rho;n_1,\ldots,n_{i})}{\prod_{a=1}^i\left(1-Q_\rho^{t^i_\ell(n_1,\ldots,n_i)}\right)}\,.\label{GeneralFormVertices}
\end{align}
Here $p_\ell^i$ are a set of $\ell$ homogeneous polynomials of order $i$ in $n_{1,\ldots,i}$ while $t^i_\ell$ are a set of $\ell$ linear functions in $n_{1,\ldots,i}$. Furthermore, the $\Lambda_\ell^i$ are rational functions in $Q_{\widehat{a}_{1,\ldots,i}}=e^{2\pi i \widehat{a}_i}$ and $Q_\rho$. They 

\begin{wrapfigure}{r}{\dimexpr 0.47\textwidth + 2\FrameSep + 2\FrameRule\relax}
\vspace{-0.6cm}
\begin{shaded*}\raggedleft
\scalebox{1}{\parbox{8.1cm}{\begin{tikzpicture}[scale = 1.50]
\draw (-0.975,1.025) circle (0.05);
\draw (-0.975,-1.025) circle (0.05);
\draw[ultra thick] (-0.95,1) -- (0,0);
\draw[ultra thick] (-0.95,-1) -- (0,0);
\draw[ultra thick] (-1.15,-0.75) -- (0,0);
\draw (-1.18,-0.77) circle (0.05);
\draw (-1.385,-0.255) circle (0.05);
\draw (-1.18,0.77) circle (0.05);
\draw (-1.385,0.255) circle (0.05);
\node[rotate=65] at (-1.2,0.5) {$\cdots$};
\draw[ultra thick] (-1.35,-0.25) -- (0,0);
\draw[ultra thick] (-1.35,0.25) -- (0,0);
\node[rotate=-65] at (-1.2,-0.5) {$\cdots$};
\draw[ultra thick] (-1.15,0.75) -- (0,0);
\node at (-1.2,1.3) {$W_{(0)}$};
\node at (-1.5,0.95) {$W_{(0)}$};
\node at (-1.75,0.325) {$W_{(0)}$};
\node[rotate=58] at (-2.4,1.2) {\parbox{2cm}{\small $(N-1-i)$ \\[-6pt]${}$\hspace{0.5cm}times}};
\node at (-1.2,-1.3) {$H_{(0)}^{(0,1)}$};
\node at (-1.6,-0.85) {$H_{(0)}^{(0,1)}$};
\node at (-1.8,-0.31) {$H_{(0)}^{(0,1)}$};
\node[rotate=-55] at (-2.2,-1.2) {\parbox{1.3cm}{\small $i$ times}};
%
\draw[ultra thick] (1.35,-0.35) -- (0,0);
\draw[ultra thick] (1.35,0.35) -- (0,0);
\draw (1.385,0.355) circle (0.05);
\draw (1.385,-0.355) circle (0.05);
\node at (1.85,0.45) {$H_{(s_1)}^{(0,1)}$};
\node[rotate=90] at (1.4,0) {$\cdots$};
\node at (1.85,-0.45) {$H_{(s_k)}^{(0,1)}$};
\draw[ultra thick,<->] (-1.5,-1.6) to [out=143,in=280] (-2.3,-0.35);
\draw[ultra thick,<->] (-1.5,1.6) to [out=-143,in=-280] (-2.3,0.35);
\draw[ultra thick,fill=black] (0,0) circle (0.05);
\draw[thick,fill=gray!60!white] (0,0) circle (0.41);
\node[rotate=0] at (0,0) {$\cpf{N}{i}_{r,\underline{s}}$};
\end{tikzpicture}}}
\caption{\sl Schematic contribution to $B^{(N,r)}_{(s)}$ as $(N-1-i)$  $W_{(0)}$ and $i$ $H_{(0)}^{(0,1)}$ coupling to multiple $H_{(s_a)}^{(0,1)}$ through $\cpf{N}{i}_{r,\underline{s}}$, for $i=0,\ldots,N-1$ and $a=1,\ldots,k=N(r-1)+1$.}
\label{Fig:SummaryNPointFctHigherOrder}
%
\end{shaded*}
\vspace{-0.7cm}
\end{wrapfigure}
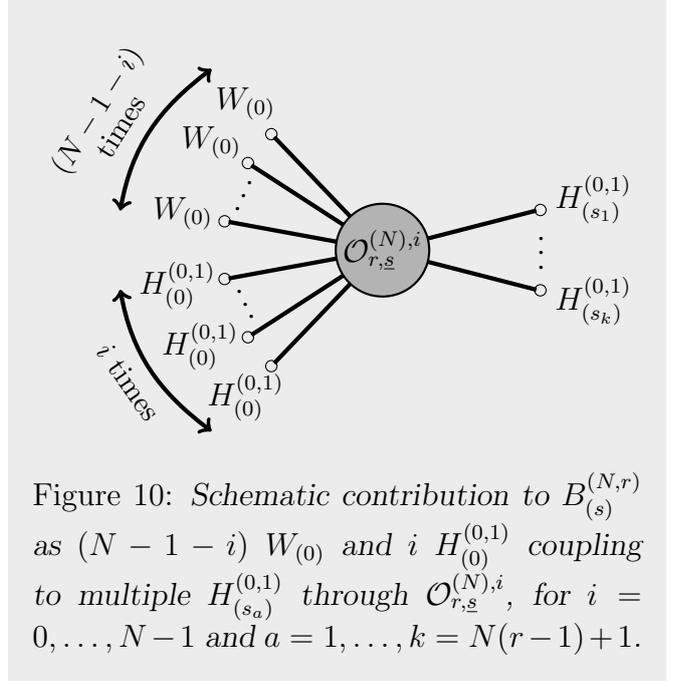

\noindent
can in general be written as combinations of sums over (parts of) the root lattice of the algebra $\widehat{\mathfrak{a}}_{N+1}$. Finally, in the notation in (\ref{GeneralFormVertices}) it is understood that $\mathcal{O}^{(N),i=0}=\text{const.}$, and in fact the examples we have studied explicitly suggest\footnote{We have also verified this for $N=4$ up to order $s=4$.}
\begin{align}
\mathcal{O}^{(N),i=0}=N\,,
\end{align}
which is in agreement with the proposed T-duality of little string theories~\cite{Hohenegger:2016eqy}. While in this work, we have only studied higher orders in $Q_R$ for $N=2$, it is already clear from this example that the former exhibits more complicated structures, generalising \figref{Fig:SummaryNPointFct} in several ways. On the one hand side, to order $Q_R^r$, we find terms that can be obtained from the free energy at order $Q_R^1$ in (\ref{GenFreeEnergyOrd1}) through the action of Hecke operators or through multiplying the arguments by suitable integers. Examples of this type in $N=2$ are given written in (\ref{N2OrdR2NonOrb}) and (\ref{N2OrdR3NonOrb}). Furthermore, we also expect to find new terms, which can schematically be represented as in \figref{Fig:SummaryNPointFctHigherOrder}. The latter correspond to coupling $(N-1-i)$ factors of $W_{(0)}$ and $i$ factors of $H_{(0)}^{(0,1)}$ (for $i=0,\ldots,N-1$) to $k=N(r-1)+1$ factors $H_{(s_a)}^{(0,1)}$ where $a=1,\ldots,k$. Here the coupling functions $\cpf{N}{i}_{r,\underline{s}}$ a priori are different for distinct choices of the external states $(H^{(0,1)}_{(s_1)},\ldots,H^{(0,1)}_{(s_k)})$. They are therefore labelled by the set of integers $\underline{s}=\{s_1,\ldots,s_k\}$. Contributions of the type shown in \figref{Fig:SummaryNPointFctHigherOrder} for $N=2$ are exhibited in (\ref{CouplingN2r2cpf}) for $r=2$ and (\ref{N2r3cpf}) for $r=3$. Since they are difficult to analyse (and we have in fact not been able to determine a general pattern), we shall content ourselves with the schematic representation in \figref{Fig:SummaryNPointFctHigherOrder} and shall not discuss them further in this work.

\subsection{Modular Graph Functions}\label{Sect:GraphFunctions}
Throughout the computations in Sections~\ref{Sect:ExN2} and \ref{Sect:ExN3} as well as in the previous subsection we have invoked a graphical representation for instanton contributions to the little string free energy that resembles higher-point functions representing effective couplings. While primarily a useful graphical device to organise the different contributions, we have already remarked previously that there might be more to it, beyond a simple graphical similarity. In this section, focusing mostly on the order $Q_R^1$, we provide further evidence to this effect.

\subsubsection{Propagator}\label{Sect:PropN2}
We start by considering the simplest (non-trivial) coupling function $\cpf{N}{1}$, for which we have found the expressions in eq.~(\ref{Sect:PropN2OrderE0}) for $N=2$ and eq.~(\ref{N3StructureCouplingNonOrba}) for $N=3$. Following \cite{Paper2}, these two functions can be written in terms of the Weierstrass elliptic function $\wp(z;\rho)$ (see \ref{DefWeierstrass})
\begin{align}
\cpf{2}{1}(\widehat{a}_1,\rho)&=-\frac{2}{(2\pi i)^2}\left[\frac{\pi^2}{3}\,E_2(\rho)+\wp(\widehat{a}_1;\rho)\right]\,,\label{N2SimpPropRev}\\
\cpf{3}{1}(\widehat{a}_1,\widehat{a}_2,\rho)&=\frac{2}{(2\pi i)^2}\sum_{\ell=1}^3\left[\frac{\pi^2}{3}\,E_2(\rho)+\wp(\widehat{a}_\ell;\rho)\right]\,,&&\text{with} &&\widehat{a}_3=\rho-\widehat{a}_1-\widehat{a}_2\,.\label{N3SimpPropRev}
\end{align}
Written in this form, however, we can express $\cpf{2}{1}$ and $\cpf{3}{1}$ in terms of the two-point function of a free boson on the torus (see appendix~\ref{App:TorusPropagator})
\begin{align}
\mathbb{G}(z;\rho)=-\ln\left|\frac{\theta_1(z;\rho)}{\theta'_1(0,\rho)}\right|^2-\frac{\pi}{2\text{Im}(\rho)}\,(z-\bar{z})^2\,.\label{FormPropagator}
\end{align}
Following \cite{Eguchi:1986sb,Kuzenko:1991vu} the latter satisfies $\partial_z^2\,
\mathbb{G}(z;\rho)=\mathbb{G}''(z;\rho)=\wp(z;\rho)+\frac{\pi^2}{3}\,\widehat{E}_2(\rho)$ (where $\widehat{E}_2(\rho)$ is defined in (\ref{NonholEisensteinD})), such that we can write
{\allowdisplaybreaks
\begin{align}
\cpf{2}{1}(\widehat{a}_1,\rho)&=-\frac{2}{(2\pi i)^2}\left[\mathbb{G}''(\widehat{a}_1;\rho)+\frac{2\pi i}{\rho-\bar{\rho}}\right]\,,\nonumber\\
\cpf{3}{1}(\widehat{a}_1,\widehat{a}_2,\rho)&=\frac{2}{(2\pi i)^2}\sum_{\ell=1}^3\left[\mathbb{G}''(\widehat{a}_\ell;\rho)+\frac{2\pi i}{\rho-\bar{\rho}}\right]\,,&&\text{with} &&\widehat{a}_3=\rho-\widehat{a}_1-\widehat{a}_2\,.\label{IdentityTorusTwoPoint}
\end{align}}
This makes it clear that $\cpf{2}{1}(\widehat{a}_1,\rho)$ and $\cpf{3}{1}(\widehat{a}_1,\widehat{a}_2,\rho)$ are modular objects, up to a term proportional to $(\rho-\bar{\rho})^{-1}$. Since the latter is precisely the contribution to complete the $E_2$ in (\ref{N2SimpPropRev}) and (\ref{N3SimpPropRev}) into $\widehat{E}_2$, $\cpf{N}{1}$ are (non-holomorphic) modular objects, provided $E_2$ is replaced by $\widehat{E}_2$. A similar behaviour under modular transformations has been observed previously (see \emph{e.g.} \cite{Hohenegger:2015cba}) in certain expansions of the free energy. 

Furthermore, using the data provided in \cite{Paper2}, we can compute in the same manner $\cpf{4}{1}$ (\emph{i.e.} for $N=4$), for which we find
\begin{align}
\cpf{4}{1}(\widehat{a}_1,\widehat{a}_2,\widehat{a}_3,\rho)&=-\frac{2}{(2\pi i)^2}\bigg(\frac{12\pi i}{\rho-\bar{\rho}}+\sum_{\ell=1}^4\mathbb{G}''(\widehat{a}_\ell;\rho)+\mathbb{G}''(\widehat{a}_1+\widehat{a}_2;\rho)+\mathbb{G}''(\widehat{a}_2+\widehat{a}_3;\rho)\bigg)\,,\label{N4PropagatorGen}
\end{align}
where $\widehat{a}_4=\rho-\widehat{a}_1-\widehat{a}_2-\widehat{a}_3$. We remark that the specific form of the last two terms in the bracket of eq.~(\ref{N4PropagatorGen}) is due to the fact that we are considering the free energy in which single-particle states have been removed.\footnote{This is due to the presence of the plethystic logarithm rather than the simple logarithm in (\ref{PlethLog}).} Including the latter in the full free energy for $N=4$ would lead to a more symmetric combination of arguments for the latter terms. 

\subsubsection{Modular Graph Functions}

The coupling functions $\cpf{N}{i}$ for $i>1$ are somewhat more involved. However, they still share a fair amount of properties with so-called graph functions, which have appeared in the study of Feynman diagrams in field theory as well as string theory \cite{Broedel:2015hia,DHoker:2015wxz,DHoker:2016mwo,DHoker:2017pvk,Zerbini:2018sox,Zerbini:2018hgs,Gerken:2018jrq,Mafra:2019ddf,Mafra:2019xms,Gerken:2019cxz}. To make contact with these recent results in the literature, we choose to represent (\ref{N2SimpPropRev}) in a slightly different fashion \cite{Paper2}
\begin{align}
&\cpf{2}{1}=-2\,\mathcal{I}_0\,,&&\text{with}&&\mathcal{I}_0=\sum_{n=1}^\infty\frac{n}{1-Q_\rho^n}\left(Q_{\widehat{a}_1}^n+\frac{Q_\rho^n}{Q_{\widehat{a}_1}^n}\right)\,.
\end{align}
The function $\mathcal{I}_0$ can (formally) be writen as a second derivative $\mathcal{I}_{0}=D_{\widehat{a}_1}^2\,\mathcal{I}_{-1}$ (where $D_{\widehat{a}_1}=\frac{1}{2\pi i}\,\frac{\partial}{\partial\widehat{a}_1}=Q_{\widehat{a}_1}\frac{\partial}{\partial Q_{\widehat{a}_1}}$) of the function
{\allowdisplaybreaks
\begin{align}
\mathcal{I}_{-1}&=\sum_{n=1}^\infty\frac{n^{-1}}{1-Q_\rho^n}\left(Q_{\widehat{a}_1}^n+\frac{Q_\rho^n}{Q_{\widehat{a}_1}^n}\right)=\sum_{n=1}^\infty Q^n_{\widehat{a}_1}+\sum_{n=1}^\infty\sum_{k=1}^\infty k^{-1}\,Q_\rho^{nk}\,\left(Q_{\widehat{a}_1}^k+Q_{\widehat{a}_1}^{-k}\right)\nonumber\\
&=\sum_{n=0}^\infty\text{Li}_1(Q_\rho^n\,Q_{\widehat{a}_1})+\sum_{n=1}^\infty\text{Li}_1(Q_\rho^n\,Q_{\widehat{a}_1}^{-1})\,.\label{PolyLogSimpleFirst}
\end{align}}
Comparing with the notation introduced in \cite{DHoker:2015wxz}, we have
\begin{align}
2\text{Re}(\mathcal{I}_{-1})=D_{1,1}(Q_\rho,Q_{\widehat{a}_1})-\frac{\pi\text{Im}(\rho)}{3}\,,
\end{align}
where we set $\zeta=Q_{\widehat{a}_1}$ and $u=0$. The object $D_{1,1}$ is one of the fundamental building blocks in the construction of the modular graph functions discussed in \cite{DHoker:2015wxz}. In particular, it was argued in \cite{DHoker:2015wxz} that the $I_\alpha:=(Q_{\widehat{a}_1}\frac{\partial}{\partial Q_{\widehat{a}_1}})^{2\alpha} \mathcal{I}_0$ (for $\alpha\in\mathbb{N}$), which generically appear in the expansion of the free energy, are combinations of polylogarithms, thus generalising (\ref{PolyLogSimpleFirst}). In fact, following \cite{Paper2}, the $\cpf{N}{i}$ have very similar properties: they can be written as combinations of generating functions of multiple divisor sums $T(\widehat{a}_1,\ldots,\widehat{a}_{N-1};\rho)$, which have first been introduced in \cite{Bachmann:2013wba} (see appendix~\ref{App:MultiDivisorSums} for a short overview). For example, the simplest case $\cpf{2}{1}$ in (\ref{Sect:PropN2OrderE0}) can be expressed as
\begin{align}
\cpf{2}{1}(\widehat{a}_1,\rho)=-2\,D_{\widehat{a}_1}\left[T(\widehat{a}_1-\rho;\rho)-T(-\widehat{a}_1;\rho)\right]\,.
\end{align}
In a similar fashion also other $\cpf{N}{i}$ can be represented in terms of the generating functions of the multiple divisor sums $T(\widehat{a}_1,\ldots,\widehat{a}_{N-1};\rho)$ (see \cite{Paper2}). Using the definition (\ref{DefTBachmann}) along with the presentation (\ref{PolylogExpansion}) for the latter, we see that $\cpf{N}{i}$ can also be decomposed into polylogarithms. The main difference, however, is that the latter have in general negative level. It would nonetheless be interesting to see, if these objects still allow a presentation in terms of Eichler integrals as for example in \cite{DHoker:2015wxz}. We leave this question for future work.
\section{Conclusions}\label{Sect:Conclusions}
In this paper we have continued the study of (non-perturbative) symmetries in a class of little string theories of A-type (see \cite{Paper1,Paper2}). We have focused on those theories, which in the low energy limit describe a six-dimensional gauge theory with gauge group $U(N)$ (for $N\in\mathbb{N}$) and matter in the adjoint representation. Making use of recently discovered \cite{Paper1,Paper2} patterns in the series expansion of the free energy (which we also verified to even higher order in this work), we have organised the unrefined limit of the latter (which corresponds to the choice $\epsilon_1=-\epsilon_2$ of the deformation parameters) in a rather intriguing fashion: for $N=2,3$ and to leading order in the instanton parameter (\emph{i.e.} $Q_R^1$), we have shown that the $B^{2,r}_{(s)}$ can be organised in a way that resembles sums of higher-point functions (almost like Feynman diagrams). Indeed, a general such contribution consists of $N$ 'external legs' out of which one is $H_{(s)}^{(0,1)}$ and the remaining $N-1$ are either $H_{(0)}^{(0,1)}$ or $W_{(0)}=\tfrac{1}{24}(\phi_{0,1}+2E_2\,\phi_{-2,1})$. These are either basic building blocks of the free energy for $N=1$ or related to the function $W$ that governs the BPS-counting of an M5-brane with a single M2-brane attached to it on either side (see \cite{Paper1}). These external legs are coupled through the coupling functions $\cpf{N}{i}$ (for $i=0,\ldots,N-1$). The latter are functions of the roots $\widehat{a}_{1,\ldots, N-1}$ of the gauge algebra $\mathfrak{a}_{N+1}$ as well as $\rho$, but are independent of $S$ and the deformation parameter $s$: in the simplest case, \emph{i.e.} for $i=0$, the examples we have studied (including the case $N=4$) suggest that $\cpf{N}{0}=N$ is a simple constant. For $i=1$, $\cpf{N}{i}$ is a non-trivial function and we have seen (for $N=2,3,4$) that it can be related to the second derivative of the scalar Greens function on the torus, \emph{i.e.} the scalar propagator. It is this fact which leads us to believe that the diagrammatic representation is more than a mere graphical device. Indeed, higher coupling functions $\cpf{N}{i>1}$ show certain similarities with graph functions that have been studied in the literature (see notably \cite{DHoker:2015wxz}) in connection with scattering amplitudes in string and field theory.  Higher orders in the instanton parameter show similar structures, however, they are complicated by two facts: \emph{(i)} these 'diagrams' contain additional legs of the form $H_{(s)}^{(0,1)}$ with more complicated coupling functions; \emph{(ii)} to this order we also find contributions that are obtained from the leading instanton result through the action of (Hecke) operators.

The results obtained in this work lend themselves to direct generalisations in a number of directions: on the one hand side it is interesting to understand if certain similarities between the coupling functions $\cpf{N}{i}$ and graph functions is merely a coincidence or can be made more concrete. In this regard it might be interesting whether the algebra of the generating functions of multiple divisor sums \cite{Bachmann:2013wba} leads to an algebra of the $\cpf{N}{i}$ which is akin to recent results in the amplitude literature (see \emph{e.g.} \cite{Mafra:2019ddf,Mafra:2019xms}). On the other hand, the appearance of contributions to higher order in the instanton parameters that can be obtained through the action of (Hecke) operators on the contributions to order $Q_R^1$, is very reminiscent of the Hecke like-structures observed in \cite{Ahmed:2017hfr} in a particular subsector of the free energy in the NS-limit (see also similar observations in \cite{Paper2}). It will be interesting in the future to see if it is possible to make this connection more concrete. 

\section*{Acknowledgements}
I would like to thank Brice Bastian for collaboration on the previous projects \cite{Paper1,Paper2}, which have been a great influence for this work. Furthermore, I am deeply indebted to Amer Iqbal, Oliver Schlotterer and Pierre Vanhove for a careful reading of the manuscript and many useful comments and exchanges. 
\appendix
\section{Modular Toolbox}\label{App:ModularStuff}
In an attempt to keep this work self-contained, in this appendix we compile a minimum of relations and properties regarding modular forms, which are necessary for the discussion in the main body of this article. For more information, we refer the reader to \emph{e.g.} \cite{EichlerZagier}.
\subsection{Jacobi Forms and Eisenstein Series}
In the main body of this paper we frequently encounter two sets of functions, namely the Jacobi forms $\phi_{-2,1}$ and $\phi_{0,1}$ as well as the Eisenstein series $E_{2k}$. The former are defined as 
\begin{align}
&\phi_{0,1}(\rho,z)=8\sum_{a=2}^4\frac{\theta_a^2(z;\rho)}{\theta_a^2(0,\rho)}\,,&&\text{and}&&\phi_{-2,1}(\rho,z)=\frac{\theta_1^2(z;\rho)}{\eta^6(\rho)}\,,\label{DefPhiFuncts}
\end{align}
where $\theta_{a=1,2,3,4}(z;\rho)$ are the Jacobi theta functions and $\eta(\rho)$ is the Dedekind eta function. These two are examples of weak Jacobi forms of index 1 and weight $0$ and $-2$ respectively. Under a weak Jacobi form of index $m\in\mathbb{Z}$ and weight $w\in\mathbb{Z}$ for a finite index subgroup $\Gamma\subset SL(2,\mathbb{Z})$, we understand a holomorphic function (here $\mathbb{H}$ is the upper complex plane)
\begin{align}
\phi:\,\,\mathbb{H}\times\mathbb{C}&\longrightarrow\mathbb{C}\,\nonumber\\
(\rho,z)&\longmapsto \phi(\rho;z)\,,
\end{align}
with the properties
\begin{align}
\phi\left(\frac{a\rho+b}{c\rho+d};\frac{z}{c\rho+d}\right)&=(c\rho+d)^w\,e^{\frac{2\pi i m c z^2}{c\tau+d}}\,\phi(\tau;z)\,,&&\forall\,\left(\begin{array}{cc}a & b \\ c & d\end{array}\right)\in\Gamma\,,\nonumber\\
\phi(\rho;z+\ell_1 \rho+\ell_2)&=e^{-2\pi i m(\ell_1^2\rho+2\ell_1 z)}\,\phi(\rho;z)\,,&&\forall\,\ell_{1,2}\in\mathbb{N}\,,\label{JacobiFormGen}
\end{align}
which furthermore affords a Fourier expansion of the form
\begin{align}
\phi(z,\rho)=\sum_{n= 0}^\infty\sum_{\ell\in\mathbb{Z}}c(n,\ell)\,Q_\rho^n\,e^{2\pi i z \ell}\,.
\end{align}
We can construct new Jacobi forms as polynomials in $\phi_{-2,1}$ and $\phi_{0,1}$ whose coefficients are given by suitable modular forms. In the case of the full modular group, the latter are generate by the ring of Eisenstein series spanned by $\{E_4,E_6\}$, where
\begin{align}
&E_{2k}(\rho)=1-\frac{4k}{B_{2k}}\sum_{n=1}^\infty \sigma_{2k-1}(n)\,Q_\rho^n\,,&&\forall\,k\in\mathbb{N}\,,\label{DefEisenStein}
\end{align}
and $B_{2k}$ are the Bernoulli numbers, while $\sigma_k(n)$ is the divisor sum. For certain applications, we also defined
\begin{align}
G_{2k}(\rho)=2\zeta(2k)+2\frac{(2\pi i)^{2k}}{(2k-1)!}\sum_{n=1}^\infty\sigma_{2k-1}(n)\,Q_\rho^n=2\zeta(2k)E_{2k}(\rho)\,,\label{NormEisenstein}
\end{align}
which differs by its normalisation. Notice that (\ref{DefEisenStein}) includes the case $k=1$, where $E_2$ is strictly speaking not a modular form. Instead it transforms with an additional shift term under modular transformations. It can, however, be completed into $\widehat{E}_2$, which transforms with weight $2$ under modular transformations, ate the expense of being no longer holomorphic
\begin{align}
\widehat{E}_2(\rho,\bar{\rho})=E_2(\rho)-\frac{6i}{\pi(\rho-\bar{\rho})}\,.\label{NonholEisensteinD}
\end{align}
With the help of the (holomorphic) Eisenstein series we can also define the Weierstrass elliptic function
\begin{align}
\wp(z;\rho)=\frac{1}{z^2}+\sum_{k=1}^\infty(2k+1)G_{2k+2}(\rho)\,z^{2k}\,.\label{DefWeierstrass}
\end{align}
Furthermore, we also introduce the following objects
\begin{align}
&\psi_2(\rho)=\theta_3^4(\rho,0)+\theta_4^4(\rho,0)=-2(E_2(\rho)-2E_2(2\rho))\,,&&\text{and}&&\psi_3(\rho)=E_2-3\,E_2(3\rho)\,,
\end{align}
which appear in the free energy for $N=2$ to orders $Q_R^2$ and $Q_R^3$ respectively (see \cite{Paper2}). These functions are in fact (proportional to) Eisenstein series of the congruence subgroup $\Gamma_0(2)$ and $\Gamma_0(3)$, respectively (see \cite{Lang,Stein}; see also \cite{Gaberdiel:2010ca} for a review of modular forms for congruence subgroups of $SL(2,\mathbb{Z})$ of the type $\Gamma_0(N)$). 

Finally, in the main body of this work, we shall also make use of Hecke operators. The latter map Jacobi forms of index $m$ and weight $w$ into Jacobi forms of index $km$ and weight $w$ for $k\in\mathbb{N}$. Specifically, let $J_{w,m}$ be the space of Jacobi forms of index $m$ and weight $w$ of $SL(2,\mathbb{Z})$, then the $k$th Hecke operator is defined as
\begin{align}
\mathcal{H}_k:\,\,J_{w,m}(\Gamma)&\longrightarrow J_{w,km}(\Gamma)\nonumber\\
\phi(\rho,z)&\longmapsto \mathcal{H}_k(\phi(\rho,z)) =k^{w-1}\sum_{{d|k}\atop{b\text{ mod }d}}d^{-w}\,\phi\left(\frac{k\rho+bd}{d^2},\frac{kz}{d}\right)\,.\label{DefHeckeGeneric}
\end{align}

\subsection{Torus Propagator}\label{App:TorusPropagator}
In this appendix, we provide some more background material on the scalar Greens function $\mathbb{G}(z)$, which appeared in the coupling functions $\cpf{N}{1}$ for $N=2,3,4$ (see (\ref{IdentityTorusTwoPoint}) and (\ref{N4PropagatorGen})). Our discussion follows mainly \cite{Eguchi:1986sb,Polchinski,DHoker:1988pdl}. 

Consider a free scalar field theory (with field $\phi$) on a torus whose periods of the two non-contractible cycles are chosen to be $1$ and $\rho$ respectively. This theory is invariant under differentiable reparametrisations of the torus, local Lorentz rotations and Weyl symmetry, which gives rise to numerous Ward identitites that strongly constrain the higher-point correlation functions. The Greens function is given by the two-point function
\begin{align}
\mathbb{G}(z-w;\rho)=\langle\phi(z)\,\phi(w)\rangle=-\ln\left|\frac{\theta_1(z-w;\rho)}{\theta'_1(0,\rho)}\right|^2-\frac{\pi}{2\text{Im}(\rho)}\,(z-w-(\bar{z}-\bar{w}))^2\,.
\end{align}
The latter satisfies the relation (with $\Delta$ the two-dimensional Laplacian)
\begin{align}
\Delta_z\,G(z;\rho)=4\pi \,\delta^{(2)}(z)-\frac{2\pi}{\text{Im}(\rho)}\,.
\end{align}
\section{Generating Functions of Multiple Divisor Sums}\label{App:MultiDivisorSums}
In \cite{Bachmann:2013wba} the following objects have been introduced
\begin{align}
T(X_1,\ldots,X_\ell;\rho)=\sum_{s_1,\ldots,s_{\ell}>0}[s_1,\ldots,s_{N-1};\rho]\,X^{s_1-1}\ldots X^{s_{N-1}-1}\,,\label{DefTBachmann}
\end{align}
which are generating functions of the following brackets of length $\ell$
\begin{align}
[s_1,\ldots,s_{\ell};\rho]=\frac{1}{(s_1-1)!\ldots (s_\ell-1)!}\sum_{n>0}Q_\rho^n\sum_{{u_1v_1+\ldots u_\ell v_\ell=n}\atop u_1>\ldots >u_1>0} v_1^{s_1-1}\ldots v_\ell^{s_\ell-1}\,.
\end{align}
These in turn generate multiple divisor sums
\begin{align}
&\sigma_{r_1,\ldots,r_\ell}(n)=\sum_{{u_1v_1+\ldots +u_\ell v_\ell=n}\atop{u_1>\ldots>u_\ell>0}}v_1^{r_1}\ldots v_\ell^{r_\ell}\,,&&\text{for} &&\begin{array}{l}r_1,\ldots,r_\ell\in \mathbb{N}\cup \{0\} \\ \ell,n\in\mathbb{N}\end{array}\label{DefMultiDivisor}
\end{align}
which are generalisations of the usual divisor sigma $\sigma_k(n)$, which for example appears in the definition of the Eisenstein series (see (\ref{DefEisenStein})). The brackets of length $\ell$ can also be written in the form \cite{Bachmann:2013wba}
\begin{align}
[s_1,\ldots,s_{\ell};\rho]=\sum_{n_1.\ldots >n_\ell>0}\widetilde{\text{Li}}_{1-s_1}(Q_\rho^{n_1})\ldots \widetilde{\text{Li}}_{1-s_\ell}(Q_\rho^{n_\ell})\,,\label{PolylogExpansion}
\end{align}
where the normalised polylogarithms 
\begin{align}
&\widetilde{\text{Li}}_{1-s}(z)=\frac{\text{Li}_{1-s}(z)}{\Gamma(s)}\,,&&\text{with}&& \text{Li}_{-s}=\sum_{n>0}n^s z^n=\frac{z P_s(z)}{(1-z)^{s+1}}\,.
\end{align}
In the last line we have assumed $|z|<1$ and we have introduced the Eulerian polynomials
\begin{align}
&P_s(X)=\sum_{n=0}^{s-1}A_{s,n}\,X^n\,,&&\text{with} &&A_{s,n}=\sum_{i=0}^n(-1)^i\left(\begin{array}{c}s+1 \\ i\end{array}\right)\,(n+1-i)^s\,.
\end{align}

\end{document}